\documentclass[gmd,manuscript]{copernicus}

\usepackage{mathtools}
\usepackage{bm}
\graphicspath{{figures/}}

\begin{document}
\nolinenumbers

\title{KELP v1.0: an agent-ready Lagrangian particle tracker for
unstructured meshes and HEALPix grids}

\Author[1][ivan.kuznetsov@awi.de]{Ivan}{Kuznetsov} %
\Author[1]{Alexey}{Androsov}
\Author[1]{Dmitrii}{Pantiukhin} 
\Author[1]{Claudia}{Wekerle} 
\Author[1]{Sergey}{Danilov}
\Author[1,2]{Thomas}{Jung}
\Author[1]{Nikolay}{Koldunov}

\affil[1]{Alfred Wegener Institute, Helmholtz Centre for Polar and Marine
Research, Bremerhaven, Germany} 
\affil[2]{Institute of Environmental Physics, University of Bremen, Bremen,
Germany}

\runningtitle{KELP v1.0: particle tracking on unstructured meshes and HEALPix grids}
\runningauthor{Kuznetsov et al.}

\received{}
\pubdiscuss{}
\revised{}
\accepted{}
\published{}

\firstpage{1}

\maketitle

\begin{abstract}
Lagrangian particle tracking follows virtual particles in the velocity
fields of an ocean model to study transport pathways, connectivity and
dispersion. Setting up such a study takes many technical steps, and
this work is increasingly done with coding agents, programs in which a
language model writes code, runs it and checks the result. We present
KELP v1.0, the Kinematic Engine for Lagrangian Particles, a tracker for
unstructured ocean-model meshes and HEALPix grids that is built to be
operated by such agents. KELP computes two- and three-dimensional
trajectories with established methods on triangular, quadrilateral and
mixed meshes, with global FESOM2 and coastal FESOM-C meshes as
examples, and on HEALPix, the equal-area grid on which kilometre-scale
digital twins distribute their output. Documentation and short task
instructions, called skills, tell an agent how to prepare releases and configurations, run calculations and process the trajectories. The records that this produces make each study repeatable. An agent prepared and ran all experiments of this paper with
these skills, and most of the preparation was completed within hours. In a controlled
evaluation, the skills raised the
completeness of study files prepared by five language models from
75--93\,\% to 88--100\,\%. KELP reproduced the particle distributions of
two other trackers on the same fields. Trajectories on HEALPix fields
converged to those on the model mesh as the pixels were refined. After
one year, the median distance between the two was 1.6~km at a pixel
spacing of 25.5~km. The dispersion of the ensemble agreed to within
2\,\% at 51~km and more closely at 25.5~km. HEALPix output can therefore be used for Lagrangian studies of
ensemble statistics when the original mesh is not available. In a 63-year calculation with 7.92 million particles released in the
Beaufort Gyre, transport from the gyre to Fram Strait took years to
decades, and the run took 3~h 36~min on one 128-core node. The number of particles is limited by the available
computer resources rather than by the code. A test tracked $10^{10}$
particles on a HEALPix mesh of 35.5 million wet surface nodes.

\end{abstract}

\introduction\label{introduction}

Lagrangian particle tracking computes the paths of virtual particles
carried by the velocity fields of an ocean model and answers questions
about where water comes from, where released material travels and how
regions are connected \citep{vanSebille2018}. In physical oceanography,
trajectories trace the exchange of water masses between ocean basins
\citep{Doos1995}, show the pathways of warm and dense water in the
Weddell Sea \citep{Nissen2024} and locate the source regions of
particles caught in sediment traps \citep{Wekerle2018}. In marine
biogeochemistry and biology, they follow the upwelling of carbon
\citep{Brady2021}, changes in dissolved inorganic carbon along
water-mass pathways \citep{Reijnders2024}, and larval dispersal and
population connectivity \citep{Cowen2006, CowenSponaugle2009,
Swearer2019}. Near the coast, they follow drifting oil and objects
\citep{Dagestad2018}, suspended matter in a glacial fjord
\citep{Neder2022} and changes in dispersion around offshore wind farms
\citep{Androsov2026}.

Ocean models increasingly compute on unstructured meshes, whose cells can
be refined near coasts, in straits and in regions of interest. The FESOM
family of sea ice--ocean models, for example, uses triangular meshes for
global simulations and mixed triangular--quadrilateral meshes for coastal
simulations \citep{Danilov2017, Androsov2019}. Kilometre-scale
simulations of Destination Earth, nextGEMS and EERIE distribute their
output on HEALPix grids
\citep{Hoffmann2023, Rackow2025, DoblasReyes2026, Ghosh2026}. HEALPix
divides the sphere into four-sided pixels of equal area and was first
developed for astronomical maps of the whole sky \citep{Gorski2005}. Its
resolution is set by a zoom level, where each level divides every pixel
into four and halves the pixel spacing, the square root of the pixel
area. These projects use it as a common grid for the output of different
models \citep{DoblasReyes2026}. Because interpolating the velocities
from the model mesh to HEALPix changes the coastline and the flow seen by
a particle, differences between trajectories on the two grids accumulate
over time. Meshes also contain up to tens of millions of nodes, and
multiyear integrations require reading terabytes of velocity output.

There are several existing trajectory frameworks built for structured and unstructured meshes. Ariane \citep{BlankeRaynaud1997} and TRACMASS \citep{Doos2013, Doos2017}
compute trajectories from transports through the faces of structured grid
cells. General offline frameworks include Parcels
\citep{LangeVanSebille2017, DelandmeterVanSebille2019} and OpenDrift
\citep{Dagestad2018}, and Parcels also reads FESOM2 output
\citep{ParcelsTeam2026a}. OceanTracker and PyLag track particles on
unstructured coastal meshes of models such as SCHISM and FVCOM
\citep{Vennell2021, Vennell2026, PyLag2020, PyLagDocs2026}. OceanTracker
also reads structured grids and splits every quadrilateral cell into two
triangles \citep{Vennell2026}. Ichthyop and
the Connectivity Modeling System are built for biophysical studies of
fish larvae and population connectivity \citep{Lett2008, Paris2013}.
FESOM output has been analysed with a tracker written for its triangular
mesh \citep{Wekerle2018, Nissen2024}, and coastal studies with FESOM-C
have used a particle tracker built for that model
\citep{Neder2022, Androsov2026}.

In this paper, we present KELP v1.0, the Kinematic Engine for Lagrangian
Particles. KELP tracks particles on unstructured meshes of triangles,
quadrilaterals or both, and on HEALPix grids. It does not regrid the
velocity fields and does not split quadrilateral cells into triangles.
It covers global and coastal applications in two and
three dimensions, with output of FESOM2 and FESOM-C serving as
examples in this paper. KELP builds on earlier particle tracking for
FESOM-C \citep{Neder2022, Androsov2026} and on established numerical
methods, which it combines in one engine. It keeps only the needed
velocity snapshots in memory and distributes large particle ensembles
over many processors and computer nodes, and tracking on both a model
mesh and its HEALPix version allows a controlled comparison of the two.

A tracking study on such grids involves many technical steps: reading
the model output, planning particle releases, choosing numerical schemes, running calculations on a cluster and processing the trajectories. This work is now increasingly done with coding agents
\citep{Pantiukhin2025, Pantiukhin2026}, programs in which a large
language model reads files, writes code, runs commands and checks the
result of each step before taking the next \citep{Yao2023}. Such agents
develop software \citep{Yang2024} and analyse Earth-science data, but on
complex tasks they can also fail without notice \citep{Zhou2026, Luo2026}.
Their results therefore depend on the software they operate: it should
give clear instructions, check the input and record what was done
\citep{Yang2024, Pantiukhin2026}. KELP is designed for this way of
working, while its documentation also serves users who work without an
agent. Its skills, short Markdown files with instructions for one task
each, tell the agent how to prepare a particle release, write a configuration,
start runs on a computer cluster and process the trajectory output. KELP
checks each configuration before a run starts, and the agent saves the
configuration, release files, input-data references, code version and
analysis scripts, so that a study can be repeated without the agent
conversation. Extensions of the code, such as an input module for a
further output format, use the same documented interfaces. All
experiments in this paper were prepared and run by a coding agent with
KELP, with the user setting each task, approving each launch and
checking the results.

Section~\ref{numerical-methods} describes the tracking method, and
Sect.~\ref{sec:software} the software and its operation by agents.
Section~\ref{sec:verification} verifies the implementation with
analytical tests, refinement of the mesh and of the time step,
forward--backward integrations and comparisons with two other trackers.
Section~\ref{sec:demonstrations} compares trajectories on a FESOM2 mesh
and on its HEALPix versions and follows six decades of releases in the
Beaufort Gyre. It also measures throughput and scaling and evaluates
agent-prepared studies. Sections~\ref{sec:discussion} and
\ref{sec:conclusions} discuss the results and present the conclusions.

\section{Tracking method}\label{model-description}\label{numerical-methods}

KELP describes every supported mesh in a standardized way, and the tracking
methods work on this common description. The velocity fields are used
on the mesh or grid on which they are stored, without regridding. Here we introduce the tracking method, whereas numerical details are given in Appendix~\ref{app:numerics}.

\subsection{Meshes, fields and coordinate frame}\label{meshes-and-fields}\label{tracking-frame}

\begin{figure*}[t]
\centering
\includegraphics[width=\textwidth]{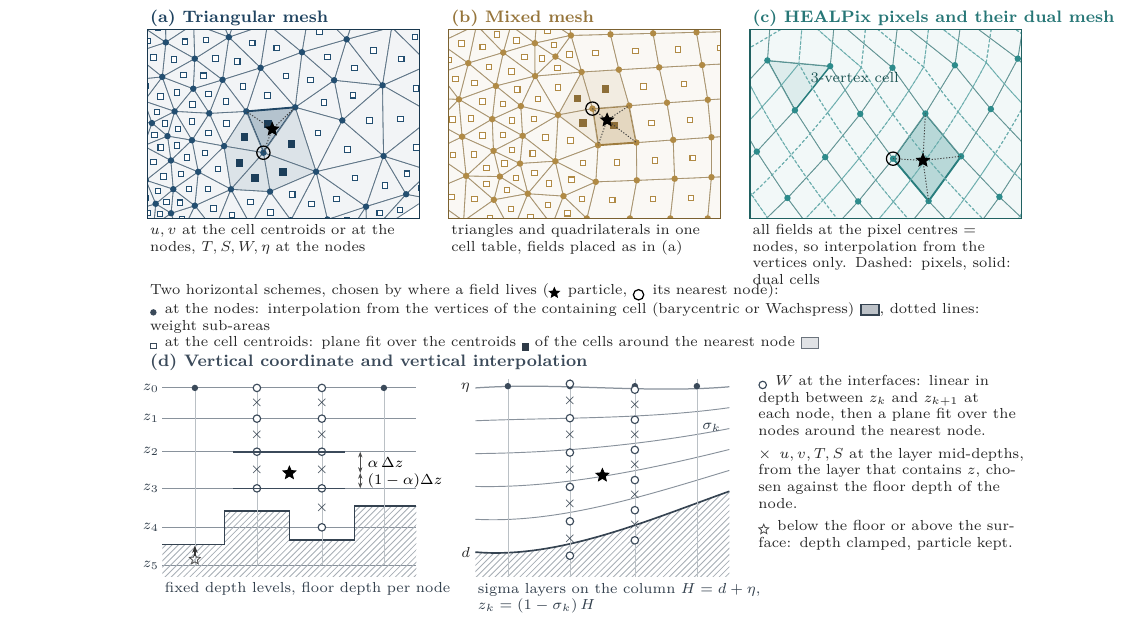}
\caption{Mesh families and the cells used for interpolation. (a) Triangular
mesh. (b) Mixed triangular--quadrilateral mesh. (c) HEALPix pixels
(dashed) and the dual mesh on their centres (solid). In (a)--(c), the
star is a particle and the open circle its nearest node. The dark cell
is the cell that contains the particle, with the sub-areas that give the
weights of its vertices. The light cells are the cells around the
nearest node, with their centroids. (d) Fixed depth levels with a floor
depth per node (left) and sigma layers (right). Open circles are the
layer interfaces, crosses the layer mid-depths and the star a
particle.}
\label{fig:meshes}
\end{figure*}

The common mesh description contains the node positions on the sphere and
the cells, which are triangles or quadrilaterals. It also lists the
nodes that form each cell and the depths of the layer interfaces
(Fig.~\ref{fig:meshes}).
Velocity and scalar fields are used where the model stores them, at the
nodes or at the cell centroids (Table~\ref{tab:hosts}). In the examples
of this paper, global FESOM2 output uses triangular cells and fixed
depth levels. Coastal FESOM-C output uses triangular or mixed
triangular--quadrilateral meshes with sigma layers.

For HEALPix output, wet pixel centres become the nodes
of a dual mesh. Each corner surrounded entirely by wet pixels defines a cell joining
their centres (Fig.~\ref{fig:meshes}c). These cells have three or four vertices. An alternative two-dimensional
representation retains the pixels as quadrilateral cells with
centroid-located velocity. Structured grids map onto the same quadrilateral
representation, but this paper does not cover them.

\begin{table*}[t]
\caption{Mesh families supported by KELP v1.0 and the positions of
their fields on the mesh. For three-dimensional fields, the horizontal
velocity and the scalars are stored at layer mid-depths and the
vertical velocity \(W\) at the layer interfaces.}
\label{tab:hosts}
\small
\newcolumntype{Y}{>{\raggedright\arraybackslash}X}
\begin{tabularx}{\textwidth}{@{}Y Y Y Y l Y@{}}
\tophline
Mesh family & Cells & Vertical coordinate & \(u, v\) & \(W\) & Scalars \\
\middlehline
Triangular (e.g.\ global FESOM2) & triangles & fixed depth levels, floor depth per node & cell centroids or nodes & nodes & nodes \\
Triangular or mixed (e.g.\ coastal FESOM-C) & triangles, quadrilaterals or both & sigma layers & cell centroids or nodes & nodes & nodes \\
HEALPix, dual representation & three- and four-vertex cells around the pixel corners & fixed depth levels & nodes (pixel centres) & nodes & nodes \\
HEALPix, pixel representation & quadrilaterals (the pixels), two dimensions only & -- & cell centroids & -- & -- \\
\bottomhline
\end{tabularx}
\end{table*}

Positions use longitude \(\lambda\) and latitude \(\phi\) in degrees,
and depth \(z\) in metres. Velocities \((U,V,W)\) are in metres per second, and
\(z\) and \(W\) are positive downward. Ocean models write \(W\)
positive upward, and KELP changes its sign when it reads the field.

An optional rotated coordinate frame places the poles on land, so that
the meridians do not converge over the Arctic Ocean. FESOM2 and HEALPix
fields can use the standard FESOM2 rotation or custom rotation angles
(Appendix~\ref{app:frame}). Release positions and output trajectories
remain in geographic coordinates.

\subsection{The time step: finding the cell and treating the boundaries}\label{one-time-step}\label{locating}\label{boundaries}

\begin{figure*}[t]
\centering
\includegraphics[width=\textwidth]{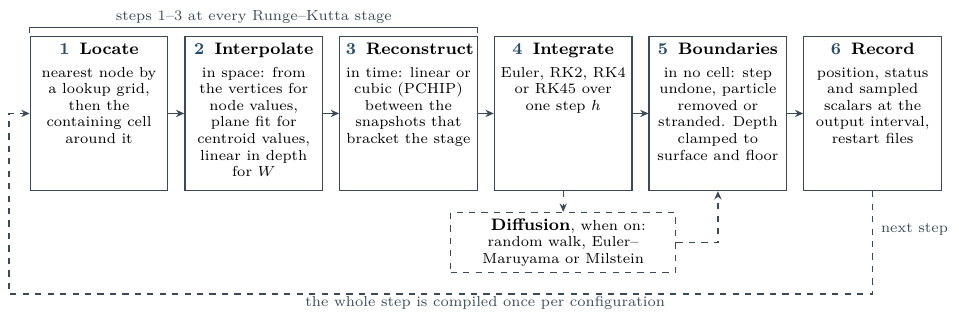}
\caption{One time step for an active particle: the six steps in order,
the bracket over steps 1--3, the optional stochastic update (dashed box)
and the return to the next step (dashed loop).}
\label{fig:step}
\end{figure*}

In every time step, each active particle passes through the same
sequence (Fig.~\ref{fig:step}). KELP first finds the cell that contains
the particle. It interpolates the velocity to the particle position in
space and then in time, advances the particle by one step and applies
the boundary rules. Diffusion adds a random displacement when it is
switched on. A particle enters this sequence at its release time. Its
position, status and optionally the sampled fields are written at the
chosen output interval.

To find the containing cell, KELP first finds the nearest node. The
nodes are sorted once into a regular longitude--latitude grid of boxes.
Only the nodes in the box of the particle and in the eight surrounding
boxes are then compared. Near the poles and the dateline, the search is widened using
distances on the sphere. The cells around the nearest node and the cells
around their vertices, two rings in all, are then tested for the
particle. An alternative is the edge walk
\citep{LohnerAmbrosiano1990, Devillers2002}. It starts from the cell of
the previous position and moves into the neighbouring cell across each
edge that the particle crossed. It stops in the cell that contains the
new position. Where the walk cannot decide, the two-ring search is
used instead (Appendix~\ref{app:localization}). The two-ring search
needs no record of the previous cell and is the default, whereas the
edge walk avoids particle losses caused by strongly stretched cells
(Sect.~\ref{sec:ver-mesh}).

When the new position of a particle lies in no cell, the step is
undone. By default, KELP removes the particle whether it crossed an open
boundary, reached a coast or was lost by the cell search. With the stranding option switched on, a particle that
reaches shallow water or a coastal cell is instead kept at its position
as stranded. It is released again when the water returns, either as soon
as the depth allows or with a given probability per step
(Appendix~\ref{app:stranding}). In the vertical, a particle is kept
between the sea surface and the sea floor at its position.

\subsection{Spatial interpolation}\label{horizontal-velocity}\label{vertical}

The horizontal interpolation depends on where the model stores the
velocity. Velocities stored at nodes are interpolated from the
vertices of the containing cell. On triangles this results in linear
(barycentric) interpolation. On quadrilaterals it uses Wachspress
coordinates, which extend bilinear interpolation to convex cells of any
shape \citep{Warren1996}. Velocities stored at the cell
centroids are interpolated with a plane fitted by least squares to the
centroid values of several cells. The cells are either those around the
nearest node, the default, or the containing cell and its edge-sharing
neighbours. A third option first averages the centroid values to the
nodes and then interpolates from the vertices. The two plane fits kept
second-order accuracy on distorted cells, whereas the averaging to the
nodes lost accuracy there (Sect.~\ref{sec:ver-mesh}). The averaged
field is continuous across cell edges, and the plane fits are not.
Appendix~\ref{app:interp} presents the cells each scheme uses and their
treatment at the coast.

Vertical velocity is interpolated linearly between the layer
interfaces at each node. The node values are then combined horizontally
with a plane fit or one of three alternatives
(Appendix~\ref{app:vertical}). In FESOM2
and FESOM-C, the vertical velocity written by the model is the flow
across the layer interfaces, which move with the sea surface. KELP
therefore uses this velocity directly to move the particle relative to
the moving layers. For FESOM2 output, KELP tracks the reference depth,
the depth that the particle would have at zero sea-surface height. It
differs from the actual depth by at most the sea-surface height.
On sigma meshes, the position relative to the layers is the fraction of
the local water column. The particle then follows the moving sea surface and the sloping
floor (Appendix~\ref{app:vertical}). Scalar fields such as temperature
are sampled at the output positions.

\subsection{Time integration and diffusion}\label{time-integration}\label{diffusion}

The fixed-step integrators are forward Euler, midpoint Runge--Kutta (RK2)
and classical fourth-order Runge--Kutta (RK4) \citep{Butcher2016}.
Adaptive integration uses the Dormand--Prince 5(4) pair
\citep{DormandPrince1980}, restricted to two-dimensional
tracking (Appendix~\ref{app:timeint}).

Temporal reconstruction is linear by default. RK2 and RK4 also accept
monotone piecewise-cubic Hermite interpolation (PCHIP)
\citep{FritschCarlson1980, FritschButland1984}.
The integration step \(h\) and snapshot interval \(\Delta T\) set distinct
errors: reducing \(h\) cannot remove the error from reconstructing stored fields.
We therefore use RK2 with linear reconstruction as the standard
setting and pair RK4 with PCHIP, because linear reconstruction would
otherwise limit RK4 to the accuracy of RK2 (Sect.~\ref{sec:ver-order}).

Horizontal diffusion uses a random walk with constant diffusivity, or
Euler--Maruyama and Milstein schemes for spatially varying diffusivity
\citep{Visser1997, RossSharples2004, Grawe2012}.
The spatially varying schemes are two-dimensional.
Vertical diffusion includes the diffusivity-gradient drift of
\citet{Visser1997} and reflects displacements at the surface and floor.
The increments and their composition with advection are given in
Appendix~\ref{app:diffusion}.

Each random number is computed directly from the particle index, the
time step and its use, such as horizontal or vertical diffusion
\citep{Salmon2011}.
A run therefore gives the same trajectories whatever the number of
threads and however the particles are divided among processes. The
same holds when a run is interrupted and restarted. Backward integration reverses
the advection while keeping the diffusion.
\section{Software and agent operation}\label{sec:software}

KELP is a Python package built on NumPy, xarray \citep{Hoyer2017} and
Numba, under the BSD 3-Clause licence. A run is described by one YAML
configuration file. It names the input data, the particle releases, the
numerical methods and the output, and KELP runs the tracking sequence
of Sect.~\ref{one-time-step} from it. This section describes how the tracking is executed on a computer cluster and how a study is set up with a coding agent.

\subsection{Execution and cost}\label{data-flow}\label{cost-model}

The velocity archive of a multiyear simulation is far larger than the
memory of a computer node. KELP therefore has an input module for each
output format. For a further output format, a coding agent can write
the input module with the same documented interface
(Sect.~\ref{sec:demo-agent}). Each module keeps only the few snapshots in memory that the current step needs and
replaces them as the run advances (Appendix~\ref{app:io}). Two options reduce the
reading cost further. Several snapshots can be read in one operation,
and the velocities can be kept in single precision, which halves their
memory. For a regional study, the fields of the region can be
extracted once from the global archive. Later runs then read only this
extract (Appendix~\ref{app:output}). Trajectories are written during
the run, and a run can stop and continue from a restart file. A long
experiment is therefore run as a chain of jobs, each within the time
limit of the computer cluster.

The work of one step is done by compiled code. For the most common
settings, the search for the cell, the interpolation and the
integration are compiled together into one routine, which avoids
intermediate arrays. Other settings run these stages one after another
(Appendix~\ref{app:kernels}). Within one process, the particles are
divided among threads. A larger ensemble is divided among processes on
one or several nodes, and their outputs are merged afterwards
(Appendix~\ref{app:output}).

The time of one step is approximately
\begin{equation}\label{eq:cost}
  t_{\mathrm{step}} \approx t_{\mathrm{I/O}} + c\,N ,
\end{equation}
where \(t_{\mathrm{I/O}}\) is the time to read the fields and \(c\) the
mean computation time per active particle, for \(N\) active particles.
Writing the trajectories adds a separate cost. The reading time does
not depend on the number of particles, so it dominates small runs on
large meshes. Both terms and their scaling are measured in
Sect.~\ref{sec:demo-performance}.

\subsection{Setting up and reproducing a study}\label{agent-layer}

\begin{figure*}[t]
\centering
\includegraphics[width=\textwidth]{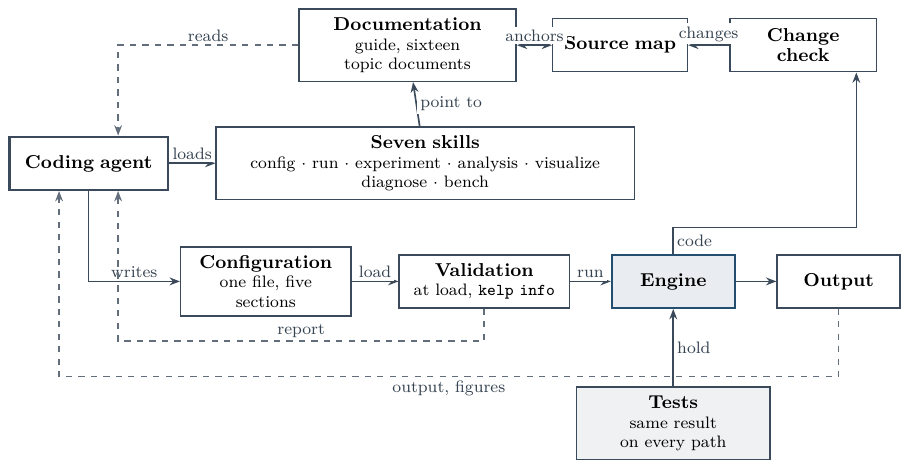}
\caption{Schematic of how a coding agent works with KELP. The agent (left) reads
the documentation (top) and the seven skills. It writes the configuration, which is checked
before the engine runs and writes the output (middle). The test suite (bottom) checks that the engine gives the same
trajectories on every execution path. Solid arrows
carry what one part gives the next, and dashed arrows return
information to the agent.}
\label{fig:agent}
\end{figure*}

We developed KELP and ran its experiments with coding agents
(Sect.~\ref{introduction}). The agent prepares the files and runs the
jobs, and the user sets the task, decides the setup and checks the
results. This shortens the routine part of a study, for example when many similar configurations are
needed for a parameter series or a scaling test. KELP includes
documentation and skills written for this working routine
(Fig.~\ref{fig:agent}).

Each skill tells the agent how to carry out one recurring task. For
example, some skills describe
how to set up a particle release and a configuration, and how to start
a run. Others describe how to crop a large mesh to a region and chain
long runs with restarts. Further skills cover counting section
crossings, comparing release cohorts and drawing figures. Each skill lists the required inputs, the steps and checks, and the
files that the task should produce. KELP provides seven skills
(Table~\ref{tab:skills}). A source map lists the functions that each
document describes, and a check reports when that code has changed. The
agent therefore does not work from an outdated description
(Appendix~\ref{app:docs}). Before a run starts, KELP checks
the configuration and the input data and refuses settings it cannot
handle (Appendix~\ref{app:config}). The user approves the prepared launch command.

The skills complement a small plotting and statistics module and
utility scripts. They guide the agent in producing the configurations,
release files, analysis scripts and figures of a particular study.
Problems and slow steps found in one campaign are written into the
skills for later studies. Section~\ref{sec:demo-agent}
reports what the agent did in the experiments of this paper and
measures what the skills add to the preparation of new studies.

A study can be repeated from its saved configuration, release files,
input-data references, code revision and analysis scripts. These
records make it independent of the conversation with the agent. The
test suite checks that a run gives the same trajectories whichever
compiled path it takes and however it is divided among processes. The
same holds for restarted runs and for runs with diffusion
(Appendix~\ref{app:tests}).

\begin{table*}[t]
\caption{The skills of KELP v1.0: the task for which an agent reads
each one and the files it produces.}
\label{tab:skills}
\centering
\begin{tabular}{lp{7.2cm}p{5.6cm}}
\tophline
Skill & Task & Output \\
\middlehline
run & Prepare a tracking run and check its input & Release file, configuration, launch command, input checks \\
config & Write, edit or check a configuration & Configuration that passes the checks at loading \\
experiment & Organize large ensembles, regional extracts and resumable campaigns & Campaign folder, code revision, jobs and result records \\
analysis & Analyse transport, release cohorts, crossings and run cost & Analysis scripts and tables \\
visualize & Figures and animations from trajectory output & Plotting scripts, figures, captions \\
diagnose & Diagnose unexpected run behaviour or cost & Diagnosis and supporting checks \\
bench & Benchmark on a given machine & Comparable performance measurements \\
\bottomhline
\end{tabular}
\end{table*}

\section{Verification}\label{sec:verification}

This section verifies that KELP computes trajectories accurately. We
proceed from tests with a known answer to tests on real model output.
In flows with an exact solution, we compare the computed trajectories
with the analytical ones (Sect.~\ref{sec:ver-analytic}). In a steady eddy on
test meshes, we measure how the error depends on the mesh spacing and
on the cell shape (Sect.~\ref{sec:ver-mesh}). In a time-varying flow,
we measure how it depends on the time step and on the interval between
the stored velocity fields (Sect.~\ref{sec:ver-order}). On ocean-model
output, where no exact trajectory exists, we track particles forward
and then backward and check how close they return to their starting positions
(Sect.~\ref{sec:ver-closure}). Finally, we compare KELP with two other
trackers by using the same velocity fields (Sect.~\ref{sec:ver-reference}).

\subsection{Analytic flows and conservation}\label{sec:ver-analytic}

Before applying KELP to ocean-model output, we verified it in simple
flows with known trajectories, as part of the automated test suite (Appendix~\ref{app:tests}). The velocity of such a flow is evaluated at the
nodes or cell centroids of a test mesh, the particles are tracked as in
a real run, and their final positions are compared with the exact ones.
A linear strain flow, a rotating flow with vertical drift and a flow
with downwelling are tracked on triangular and quadrilateral meshes and
across the interface of a mixed mesh. On the HEALPix dual mesh, uniform
zonal and meridional flows and a uniform flow with downwelling give the
exact displacements, also across the dateline, and tests near the pole
and across the dateline check the conversion between metres and degrees
(Eq.~\ref{eq:metric} in Appendix~\ref{app:timeint}). In all these
flows, the computed paths agree with the exact ones within the error of
the time step. In a steady overturning cell, each particle stays on its
streamline under each method of vertical-velocity interpolation. On a
sigma mesh with a tide over a sloping floor, Euler, RK2 and RK4 keep
the fractional depth of a particle in the moving water column exactly.

Diffusion tests check the distribution of many particles rather than
individual paths. With uniform horizontal diffusivity, the spread of the
particles grows with time at the expected rate. With a prescribed
gradient of the diffusivity, the particles drift as the theory
predicts. In a closed water column whose diffusivity increases with
depth, a uniform particle distribution stays uniform in forward and in
backward runs. The stranding rules are checked in the same way. One test dries and
rewets a flat domain with a tidal cycle, another carries a particle
onto an island shore.

\subsection{Mesh spacing and cell shape}\label{sec:ver-mesh}

Model meshes contain cells of different size and shape, in particular
near coasts. The tests of Sect.~\ref{sec:ver-analytic} show that the
schemes reproduce simple flows with exact solutions. They do not show
how the trajectory error changes with the spacing and the shape of the
cells. We measured this in a steady eddy in which every particle path is
known. Its streamfunction is $\psi = \psi_0 \exp(-r^2/2R^2)$ with
$R = 20$~km, so particles circle the eddy centre. The speed reaches
0.5~m\,s$^{-1}$ at $r = R$. One revolution takes 1.8~days at $r = 5$~km
and 40~days at $r = 50$~km. An error in the interpolated speed therefore
moves a particle ahead of or behind its exact position, and this offset
grows with time.

\begin{figure*}[tp]
\centering
\includegraphics[width=\textwidth]{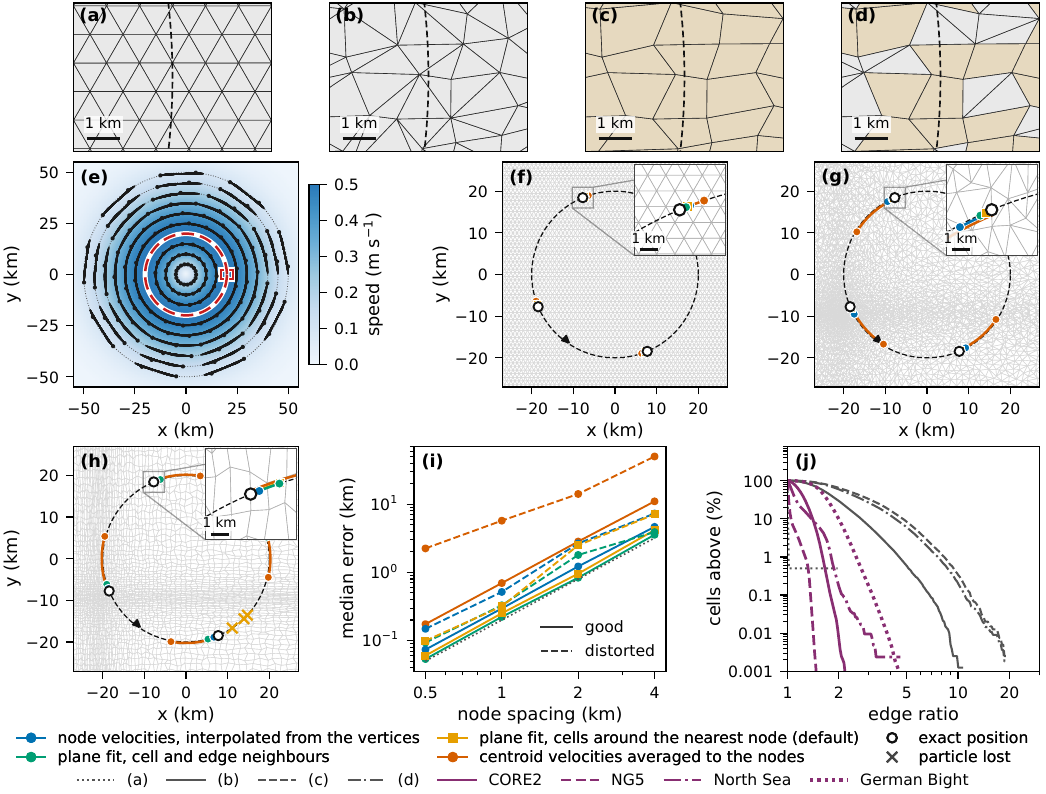}
\caption{Trajectory error in a steady eddy on test meshes with a mean
node spacing of 1~km, and cell shapes of test and model meshes.
(a)--(d) Close-ups, 6~km wide and 4.5~km high, of four test meshes:
good triangles (a), distorted triangles (b), distorted quadrilaterals
(c) and the distorted mixed mesh (d). Triangles are light grey and quadrilaterals
light sand, and the dashed line is the 20-km circle. (e) The eddy: speed (colour),
the ten release circles (dotted) and exact paths over the first 12~h
(black). The red box marks the window of (a)--(d), and the red circle the
release circle of (f)--(h). (f)--(h) Positions after 30~days of four
particles released on the 20-km circle, on the meshes of (a), (b) and
(c): exact (open circles) and computed with each scheme (colour).
Circles mark the schemes that follow the particle across cell edges
(edge walk). Squares mark the plane fit over the cells around the
nearest node (two-ring search). Crosses mark the last positions of lost particles,
and insets enlarge one particle. (i) Median error of all 720 particles after
30~days against node spacing on the good (solid) and distorted (dashed)
triangles. The dotted line has slope~2. (j) Percentage of cells whose
edge ratio (longest edge divided by shortest) exceeds a given value, for
the test meshes of (a)--(d) (grey) and four model meshes (purple):
FESOM2 CORE2, FESOM2 NG5, FESOM-C southern North Sea and FESOM-C German
Bight with wind farms. The step of (a) at 0.5~\% comes from half-size
cells along two edges of the test domain.}
\label{fig:ver-eddy}
\end{figure*}

We built three pairs of test meshes: triangles, quadrilaterals, and
triangles mixed with quadrilaterals (Fig.~\ref{fig:ver-eddy}a--d). In
each pair, one mesh is of good quality, with cells of nearly equal size and
interior angles between 30 and 90$^\circ$. The other is distorted, with
cell sizes that differ by about an order of magnitude and interior
angles from about 5$^\circ$ to 170$^\circ$. All meshes were
built with mean node spacings of 4, 2, 1 and 0.5~km. We released 720 particles on ten circles
with radii of 5--50~km and tracked them for 30~days with RK4 and a
300-s step. Velocities were sampled from the exact field at the nodes or
at the cell centroids. The error therefore comes only from the spatial
interpolation and from the search for the cell that contains the
particle. Halving the step changed the final positions by a median of
less than 2~m.

On the good-quality meshes, the error of all four schemes decreased with the
square of the node spacing, as Fig.~\ref{fig:ver-eddy}i shows for
triangles. The fitted slopes were 1.98--2.03 on triangles,
quadrilaterals and the mixed mesh, where particles also cross from one
cell type to the other. Distortion did not change this rate for the interpolation of node
velocities from the vertices or for the two plane fits (slopes
1.76--2.15). At a spacing of 0.5~km, it increased
their error by a factor of 1.4--2.2 compared with the good-quality mesh of the
same cell type. The scheme that averages centroid velocities to the nodes lost
accuracy on distorted cells. Its error decreased only about linearly
with the spacing (slopes 1.1--1.5) and was 9--18 times larger than on
the good-quality mesh. The average is exact at the area-weighted centre of the
surrounding centroids, which on distorted cells lies away from the node
(Appendix~\ref{app:interp}).

Cell shape also affected the search for the containing cell. The default
scheme, the plane fit over the cells around the nearest node, searches
two rings of cells around that node (Sect.~\ref{one-time-step}). It lost
no particles on the good-quality meshes and on the distorted triangles. On the
distorted quadrilateral and mixed meshes, it lost 24--61~\% of the
particles within 30~days (Fig.~\ref{fig:ver-eddy}h). On these cells, a
node two rows away can be closer to the particle than the corners of its
own cell. The three schemes that follow the particle across cell edges
with the edge walk lost no particles.

Real model meshes are mostly close to the good-quality test meshes, and strongly
distorted cells are rare (Fig.~\ref{fig:ver-eddy}j). We measured the
cell shapes of four model meshes in the same way as those of the test
meshes: the FESOM2 CORE2 and NG5 meshes, a FESOM-C mesh of the southern
North Sea and the FESOM-C wind-farm mesh of the German Bight used in
Sect.~\ref{sec:ver-reference}. On all four, 95\,\% of the cells have
an edge ratio (longest edge to shortest) below 2.2, and at most
0.006\,\% exceed 4, whereas on the distorted test meshes the 95th
percentile reaches 3.2--5.7 and up to 17\,\% of the cells exceed 4.
Interior angles outside 30--120$^\circ$ occur in 0--4.1\,\% of the
model cells, compared with 13--26\,\% on the distorted test meshes.
The German Bight mesh has fewer stretched cells than the distorted
triangles of Fig.~\ref{fig:ver-eddy}b, on which no scheme lost
particles. The distorted test meshes therefore represent an unfavourable
case, and where such cells occur, the schemes that use the edge walk
are the safer choice.

\subsection{Time step and snapshot interval}\label{sec:ver-order}

Offline tracking has two sources of temporal error. The integrator
approximates the trajectory within each step $h$, and the temporal
reconstruction approximates the velocity between snapshots stored every
$\Delta T$. Refining $h$ reduces only the first error.
We separate the two by varying $h$ and $\Delta T$ independently in a
flow with a known trajectory. The flow is uniform in space, so spatial
interpolation adds no error. It is an inertial oscillation superimposed
on a constant eastward current:
$U = u_g + (u_0-u_g)\cos(ft)$ and
$V = -(u_0-u_g)\sin(ft)$.
We set $f = 10^{-4}$~s$^{-1}$, $u_0 = 0.3$~m\,s$^{-1}$ and
$u_g = 0.04$~m\,s$^{-1}$, and follow one particle from the origin
for 8~h. The error is the distance of its final position from the
analytic one.

\begin{figure*}[t]
\centering
\includegraphics[width=\textwidth]{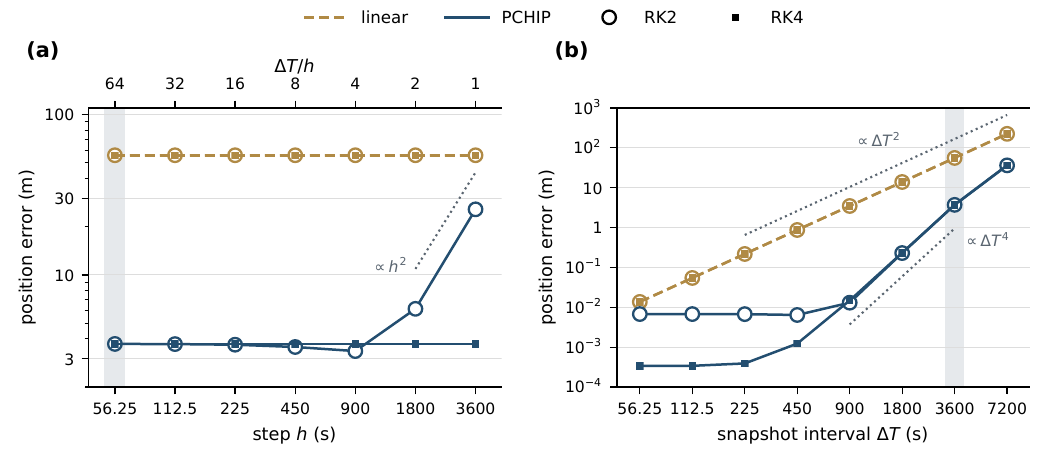}
\caption{Position error after 8~h versus (a) the integration step $h$
at $\Delta T = 1$~h, with $\Delta T/h$ on the upper axis, and (b) the
snapshot interval $\Delta T$ at $h = 56.25$~s. Dashed lines show linear reconstruction and solid lines PCHIP. Open
circles mark RK2 and filled squares RK4. Grey bands mark the runs with $h = 56.25$~s and $\Delta T = 1$~h,
shown in both panels. Dotted lines give the slopes $h^2$, $\Delta T^2$
and $\Delta T^4$.}
\label{fig:order-cap}
\end{figure*}

With linear reconstruction, the reconstruction error limits the
accuracy whatever the order of the integrator (Fig.~\ref{fig:order-cap}a).
With hourly snapshots, RK2 and RK4 gave the same 55.8-m error over the
whole range of steps.
PCHIP lowered this limit to about 3.7~m. RK4 reached it with an hourly
step, and RK2 approached it as the step was refined.

More frequent snapshots lowered the reconstruction error
(Fig.~\ref{fig:order-cap}b). At the finest step
($h = 56.25$~s, 1/64 of an hour), the error
decreased approximately as $\Delta T^2$ with linear reconstruction and
as $\Delta T^4$ with PCHIP between 15~min and 1~h.

The run skill applies these results (Table~\ref{tab:skills}): it uses
RK2 with linear reconstruction by default and pairs RK4 with PCHIP,
because RK4 gains nothing over RK2 when the reconstruction is linear.

\subsection{Forward--backward closure}\label{sec:ver-closure}\label{sec:ver-reverse}

When the exact trajectory is unknown, a forward--backward test shows
how much numerical error accumulates. A particle is tracked forward for
some time and then backward for the same time in the same fields. In
exact arithmetic it would return to its starting point. We call the
distance between the start and the final position the closure.
Diffusion is switched off, because a random walk does not retrace its
path backward. In uniform flow, the closure is zero to rounding error.

On ocean-model output, we measured the closure in KELP and in Parcels
on the same fields (Sect.~\ref{sec:ver-reference}). Three effects keep
it above zero. The integration schemes are not exactly reversible,
small differences grow where neighbouring trajectories diverge, and the
boundary rules act in one direction only. A particle removed at a
boundary cannot return, and a depth that was limited at the sea floor
cannot be restored on the way back.

\subsection{Comparison with other trackers}\label{sec:ver-reference}

Where no exact trajectory is known, we compared KELP with two other
trackers on the same velocity fields.
Each comparison used matched release positions and matched time integration.
In chaotic flow, small differences at each step separate individual
particles over time.
We therefore compare particle distributions first and the separation
of the same particle second.

The first comparison repeated the particle experiment of
\citet{Androsov2026} with KELP and with their tracker.
That study examined how offshore wind farms in the German Bight change
the tidal flow of the southeastern North Sea and the spreading of
passive particles. It used the depth-averaged version of FESOM-C,
forced only by barotropic tides at the open boundary.
Two simulations differed in the representation of 692 turbine
foundations. In the reference run, mesh cells filled the foundation
sites. In the wind-farm run, the foundations were cut out of the mesh.
Both triangular meshes have about 1.19 million nodes, with spacing
from 1.4~m around the turbines to about 1.4~km near the open boundary.
Velocities were stored at cell centroids every 300~s.

\begin{figure*}[t]
\centering
\includegraphics[width=\textwidth]{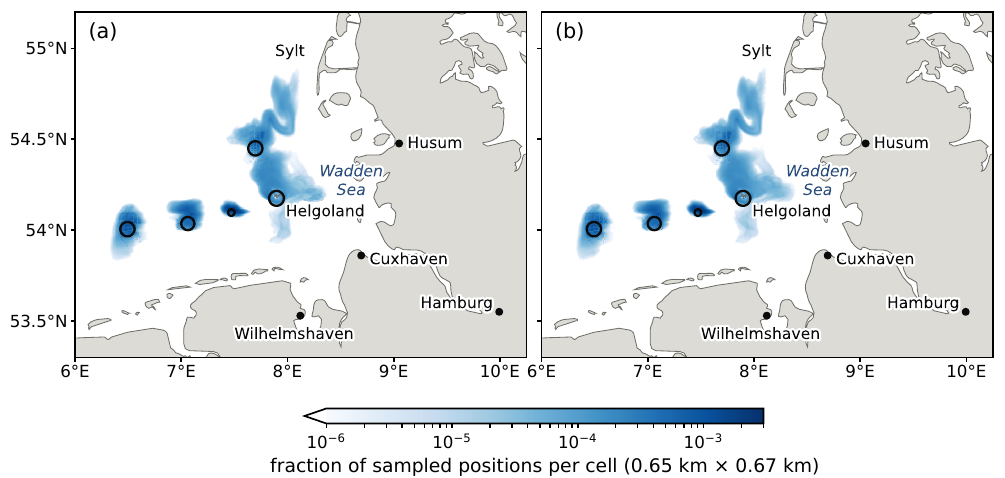}
\caption{Particle probability maps for the wind-farm run of
\citet{Androsov2026}, computed with (a) the tracker of \citet{Androsov2026} and
(b) KELP. Colours give the fraction of all sampled positions per cell of
about $0.65\times0.67$~km, with the five clouds weighted equally.
Positions were sampled over the eight weeks following each of the 78
releases. Black circles mark the release areas of the five clouds.}
\label{fig:ver-wf-maps}
\end{figure*}

The particle experiment compared the distributions of particles
released in the two runs. Five clouds of 100 particles each were
released at the same positions in both runs (Fig.~\ref{fig:ver-wf-maps}).
Three clouds lie inside wind farms, one at Helgoland and one in open
water. Each cloud was released 78 times at two-hour intervals, and each
release was followed for eight weeks. Maps of particle probability
summarise the result. They give the fraction of all sampled particle
positions that falls into each map cell.

We configured KELP with the cell search and interpolation of their
tracker. Both codes find the nearest mesh node and search the
cells around it for the particle. Both fit a least-squares plane to the
centroid velocities of these cells (Sect.~\ref{horizontal-velocity}).
Both advance particles with forward Euler steps of 300~s, equal to the
snapshot interval.
The runs of both codes were performed on a single node of the AWI
cluster Albedo.

The probability maps of the two codes overlapped by 95\,\% with wind
farms (Fig.~\ref{fig:ver-wf-maps}) and by 97\,\% without them.
The overlap is the sum over all map cells of the smaller of the two
probabilities.
For single clouds, the overlap ranged from 91 to 99\,\%.
The areas that contain 50 and 90\,\% of the probability differed by
at most 7\,\%.

After 56 days, the cloud centres of the two codes were 0.2--1.7~km
apart, while the root-mean-square radius of the clouds had grown to
1--16~km. Individual particles separated by nearly the cloud radius in two clouds
with wind farms and in one without.
The two codes therefore agree in the particle distributions, which the
experiment compares.
In the runs with their tracker, 67 of the 78 releases started
0.5--1.5~h earlier than scheduled, which likely explains part of the
difference.
Section~\ref{sec:discussion} examines the sensitivity of transport
diagnostics to finer integration steps.

The second comparison used Parcels v4 (version 4.0.1.dev7,
\citealp{ParcelsTeam2026a}) on the FESOM2 CORE2 mesh north of
50$^\circ$\,N with daily velocities of 1980. On 1 January 1980, we
released 31\,417 particles in the Beaufort Gyre between 0 and 500~m
depth and tracked them for 363 days, together with a surface set at
their 2\,250 horizontal positions. Both codes used RK4 with a 3-h step,
barycentric interpolation in the horizontal and linear interpolation in
time. Our run script supplied the vertical velocity and the time
interpolation to Parcels, whose code was not changed. These runs were
performed on a single node of DKRZ Levante (Sect.~\ref{sec:demo-performance}).

\begin{figure*}[t]
\centering
\includegraphics[width=\textwidth]{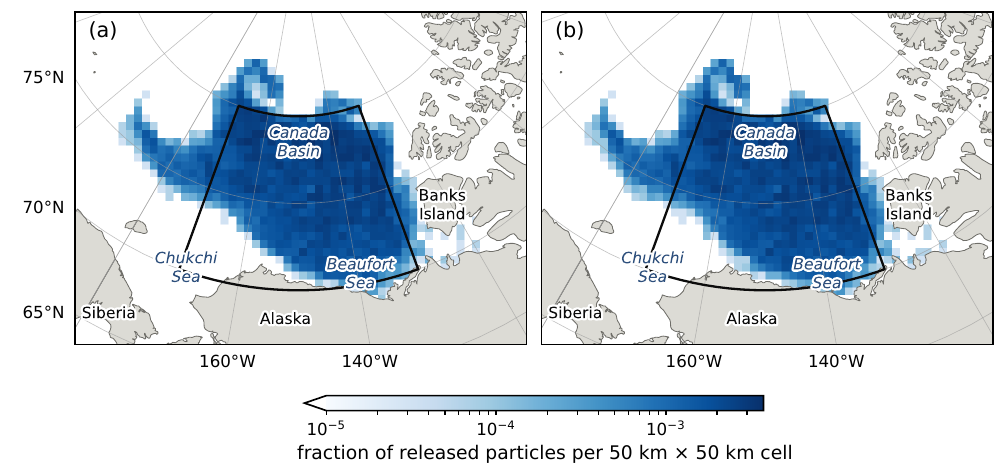}
\caption{Particle distribution one year after release, computed with
(a) KELP and (b) Parcels v4. Particles were released in the area
outlined in black. Colours give the fraction of released particles per
$50\times50$~km cell of an equal-area projection.}
\label{fig:ver-parcels-maps}
\end{figure*}

After one year, the particle distributions on $50\times50$~km cells
overlapped by 99.2\,\%, computed as for the wind-farm maps
(Fig.~\ref{fig:ver-parcels-maps}). Individual three-dimensional
trajectories separated slowly between the two codes, from a median of
8~m after 30 days to 58~m after one year, with a 90th percentile of
313~m. A 1-m shift of all release positions in KELP grew to a median of
only 1.8~m over the same year, so the separation indicates a small
systematic difference at each step rather than amplified rounding. The
largest separations occurred over shallow topography: particles
released where the bottom is shallower than 500~m formed 8\,\% of the
release but 60\,\% of those more than 10~km apart, and the 99th
percentile reached 25.6~km. In the surface set, the median separation
was 41~m and the maximum 5.7~km.

We also measured forward--backward closure in both codes
(Sect.~\ref{sec:ver-closure}, Table~\ref{tab:ver-roundtrip}), each
integrating backward for 363 days from its own final positions. Most
particles returned close to their release positions. The median closure
in 3-D was 0.1~mm in KELP and 1.6~m in Parcels, which likely reflects
the single-precision positions of the Parcels runs. More particles
failed to return in KELP than in Parcels, 1.2\,\% against 0.6\,\%
ending more than 1~km from their release positions, and in both codes
these particles lay over the shelf and slope. Most of the KELP
particles that did not return (82\,\%) had come close to the bottom on
the forward leg, so the difference likely comes from the treatment of
the bottom: KELP limits the particle depth to the water column, a step
that cannot be reversed, whereas the Parcels runs set velocities below
the local bottom to zero. The same treatment likely contributes to the
tail of the forward separations over the shelf and slope. In the
surface set, where the bottom plays no role, all KELP particles
returned, and 1.75\,\% of the Parcels particles did not.

\begin{table}[t]
\caption{Forward--backward closure for KELP and Parcels v4: number of
particles that completed both legs, median and 90th percentile of the
closure, and fraction with closure above 1~km (not returned).}
\label{tab:ver-roundtrip}
\centering
\begin{tabular}{llrrrr}
\tophline
Set & Code & Particles & Median & 90th pct. & Not returned \\
\middlehline
3-D & KELP    & 31\,357 & 0.1~mm & 1.9~cm & 1.19\,\% \\
3-D & Parcels & 31\,362 & 1.6~m  & 12~m   & 0.64\,\% \\
Surface & KELP    & 2\,231 & 0.1~mm & 7~mm & 0\,\% \\
Surface & Parcels & 2\,231 & 2.0~m  & 22~m & 1.75\,\% \\
\bottomhline
\end{tabular}
\end{table}

\section{Demonstrations}\label{sec:demonstrations}

The demonstrations test KELP across field representations, long
archives and large particle ensembles. We first track matched releases on a FESOM2 mesh
and on HEALPix versions of its fields to measure how the
representation changes
trajectories and dispersion statistics. The Beaufort Gyre experiment follows
repeated releases over six decades, using regional extraction, restart files and output written
during the run. We then measure the cost of complete
runs and the scaling with particle count, mesh size and processor count.
Finally, Sect.~\ref{sec:demo-agent} describes how a coding agent
prepared the experiments of this paper and measures what the skills add
when an agent prepares a new study.

\subsection{Effect of remapping FESOM fields to HEALPix}\label{sec:demo-healpix}

Remapping velocity fields changes the spatial interpolation and the
representation of the coastline. These differences accumulate along
a trajectory, even when the initial position and integration scheme
are unchanged. We measured their combined effect by tracking the same
release on the FESOM2 mesh and on three HEALPix versions of its
fields. The
trajectories on the FESOM2 mesh serve as the reference for this
comparison.

We used daily-mean output for 2005 from a FESOM2 simulation
\citep{Danilov2017} on the CORE2 mesh, with a nominal resolution of
about $1^\circ$ and 47 layers, forced with JRA55-do \citep{Tsujino2018}.
The node spacing of this mesh is about 25~km north of $50^\circ$N,
about 55~km near the equator and up to about 110~km in the subtropics. We remapped these fields to HEALPix grids at zoom levels 6, 7 and 8,
with pixel spacings ($\ell_\mathrm{pix}$) of 102, 51 and 25.5~km. The HEALPix fields were built by linear
interpolation on a triangulation of the model-mesh nodes, with land masked
at each depth.

The surface release contained 133\,549 particles on an equal-area
layout. A further 26\,708 particles, each 10~km east of a release
position, formed pairs. A three-dimensional
release contained 37\,773 particles at 100, 500 and 1500~m. All
four grids used the same releases and settings: interpolation from the
vertices (Sect.~\ref{horizontal-velocity}), the rotated frame, a 6-h step and
no random walk. Each grid was run twice, once with RK4 and PCHIP and once with RK2 and
linear reconstruction, to compare the effect of the temporal scheme
with that of remapping.

\begin{figure*}[t]
\centering
\includegraphics[width=\textwidth]{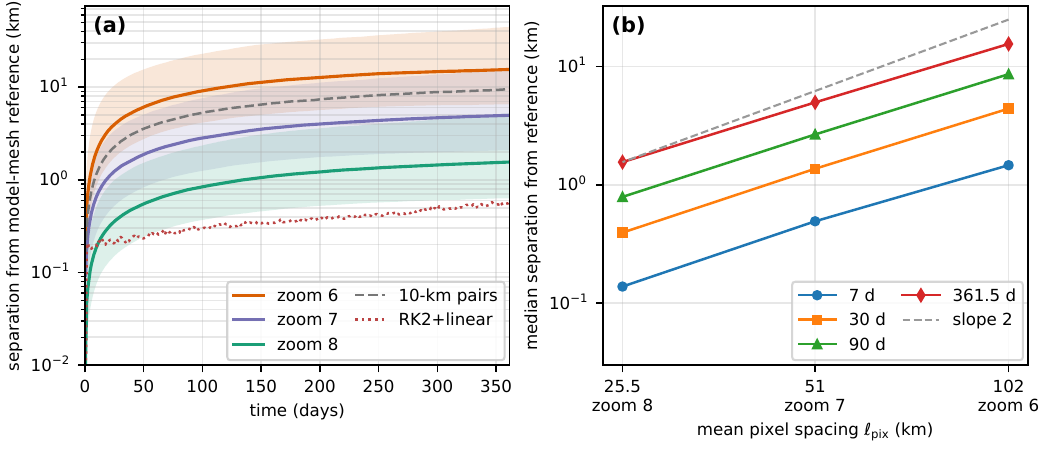}
\caption{Separation between HEALPix and model-mesh trajectories for the
surface release. (a) Median separation versus time for zooms 6--8, with
the interquartile range shaded. Grey dashed: median change of the
separation vector of model-mesh pairs released 10~km apart,
$|\mathbf{r}(t)-\mathbf{r}_0|$. Red dotted: RK4 with PCHIP
versus RK2 with linear reconstruction on the model mesh. (b) Median
separation versus $\ell_\mathrm{pix}$ at four times
after release. The grey dashed line has slope 2.}
\label{fig:e1-sep}
\end{figure*}

For each release position, we measured the distance along the sphere
between the HEALPix trajectory and the trajectory on the model mesh.
Both were computed with RK4 and PCHIP. The same measure, averaged over
trajectories, is used in meteorology to quantify the error of
trajectories computed from wind fields of reduced resolution
\citep{Stohl1995, Stohl1998}.
The median of this separation grew with time and
decreased as the HEALPix grid was refined (Fig.~\ref{fig:e1-sep}a,
Table~\ref{tab:e1-sep}). After 30 days, it was 4.4, 1.4 and 0.4~km at
zooms 6, 7 and 8. After one year it was 15, 5.0 and 1.6~km.

Each halving of $\ell_\mathrm{pix}$ reduced the median separation by a
factor of about three at all four times shown in Fig.~\ref{fig:e1-sep}b.
The separation therefore decreases approximately as
$\ell_\mathrm{pix}^{1.7}$, slightly slower than $\ell_\mathrm{pix}^{2}$. The
three-dimensional release showed the same decrease, with medians of 19,
6.5 and 2.1~km after one year (Table~\ref{tab:e1-sep}).

\begin{table}[t]
\caption{Median separation (km) between HEALPix and model-mesh (CORE2)
trajectories, using RK4 with PCHIP, for the surface (2-D) and
three-dimensional (3-D) releases.}
\label{tab:e1-sep}
\centering
\begin{tabular}{llrr}
\tophline
release & zoom ($\ell_\mathrm{pix}$, km) & 30 d & 1 year \\
\middlehline
2-D & 6 (102)  & 4.4 & 15 \\
2-D & 7 (51)   & 1.4 & 5.0 \\
2-D & 8 (25.5) & 0.4 & 1.6 \\
3-D & 6 (102)  & 3.3 & 19 \\
3-D & 7 (51)   & 1.1 & 6.5 \\
3-D & 8 (25.5) & 0.3 & 2.1 \\
\bottomhline
\end{tabular}
\end{table}

Two quantities from the model mesh alone put these separations in
scale (Fig.~\ref{fig:e1-sep}a). For particles released 10~km apart, the
vector between them changed by a median of 9.5~km after one year, about
six times the zoom-8 separation. At all times, this change lay between
the zoom-6 and zoom-7 separations.
Changing the temporal scheme from RK4 with PCHIP to RK2 with linear
reconstruction separated the trajectories by a median of 0.23~km after
30 days, and a smaller step did not reduce this difference
(Sect.~\ref{sec:ver-order}). The temporal scheme therefore mattered less
than the grid, although at zoom 8 the two effects were comparable
(0.23 and 0.39~km).

\begin{figure*}[t]
\centering
\includegraphics[width=\textwidth]{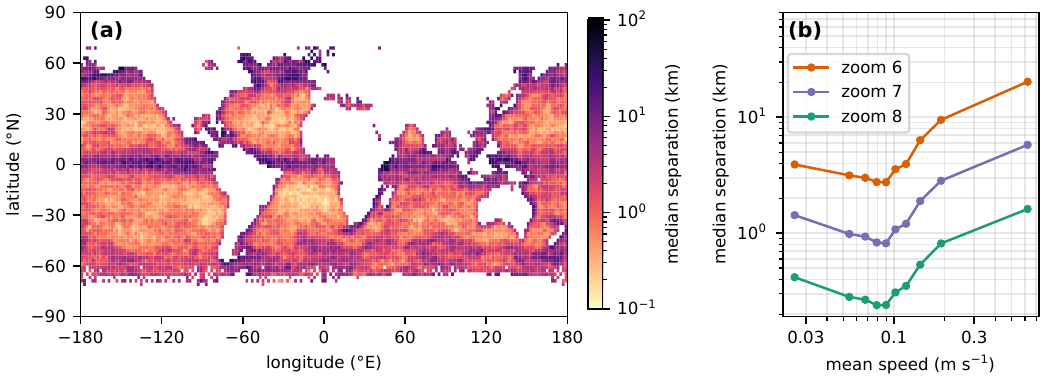}
\caption{Regional differences between HEALPix and model-mesh trajectories for the
surface release after 30 days. (a) Median separation at zoom 7,
grouped by the $2^\circ$ cell of release. Cells with fewer than eight
particles are blank. (b) Median separation for zooms 6--8, grouped by
deciles of mean speed along the model-mesh trajectory.}
\label{fig:e1-dyn}
\end{figure*}

The separation depends on the region of release. After 30 days at
zoom 7, the largest median separations followed the equatorial band,
the western boundary currents, the Antarctic Circumpolar Current and
marginal seas (Fig.~\ref{fig:e1-dyn}a). In the interiors of the
subtropical gyres, the medians stayed below 1~km.
The separation also depends on the flow speed along the trajectory.
Grouped by deciles of mean speed along the model-mesh trajectory, the zoom-7
median was smallest, 0.8~km, at 0.08--0.09~m\,s$^{-1}$ and reached
5.8~km in the fastest decile (Fig.~\ref{fig:e1-dyn}b). It also rose to
1.4~km in the slowest decile, which is dominated by near-coastal
releases. Zooms 6 and 8 show the same dependence on speed.

The spreading of the release as a whole agreed much more closely than
individual trajectories. For each particle of the surface release, we
took the squared distance from its own release position and averaged
it over all particles. This absolute dispersion measures how far the
particles have spread on average, regardless of where each one went
\citep{Taylor1922, LaCasce2008}. After one year, it was about 2\,\% below the
model-mesh value at zoom 7, and the difference decreased at finer pixel
spacing. Individual trajectories at zoom 7 were then a median 5~km
apart.
In idealized turbulence, absolute dispersion was likewise insensitive
to the spatial sampling of the velocity field \citep{Keating2011}.
Relative dispersion was sensitive to this sampling in one of their two
flows.

\subsection{Six decades of releases in the Beaufort Gyre}\label{sec:demo-arctic}

The Beaufort Gyre is the largest freshwater reservoir of the Arctic
Ocean \citep{Proshutinsky2019, TimmermansToole2023}. It accumulates
freshwater under persistent anticyclonic winds and releases it when the
winds weaken, over years to a decade \citep{TimmermansToole2023}.
Released freshwater leaves the Arctic on both sides of Greenland,
through the Canadian Arctic Archipelago and Davis Strait and through
Fram Strait \citep{Zhang2021, Wang2024}. In the subpolar North Atlantic,
it freshens the upper ocean and can affect stratification and deep
convection \citep{Zhang2021, TimmermansToole2023}. In model experiments
with a passive tracer, released freshwater took several years to reach
these gateways \citep{Wang2024}. A large part of it stayed in the Arctic
for many years.

A tracer, however, does not keep the release time and depth of the water
once it mixes. Lagrangian particles keep them and show the routes,
depths and transit times of this water directly \citep{Lique2010}. We
tracked particles released in the gyre every three months over 63
years, from 1958 to 2020. Particles released in different decades were
compared at the same age, ten years after release, and at their passage
through Fram Strait.

We used daily velocity fields from the CORE2 simulation described in
Sect.~\ref{sec:demo-healpix}. The tracking domain is the mesh north
of $50^\circ$N, with a resolution of about 25~km,
49\,154 nodes and all 47 layers. It was extracted from the global
output once, before tracking. Every three months, we released
31\,417 particles in the Beaufort Gyre box,
130--170$^\circ$W and 70--80$^\circ$N. The releases covered four depth
bands: 0--50, 50--100, 100--250 and 250--500~m. The 252 releases
contained 7.92 million particles in total. The tracker wrote particle
positions every five days.

Particles were advected by the model velocities alone, without a
random walk or stranding. We used RK2 with linear temporal
reconstruction and a one-day step, equal to the snapshot interval.
Horizontal velocities at the nodes were interpolated barycentrically,
and vertical velocity with a least-squares plane fit.

For each depth band, the 40 releases of one decade form a cohort of
0.29 to 0.34 million particles. We compared cohorts from the 1970s to
the 2000s using each particle's position ten years after release.
All statistics of a cohort are fractions of the number of particles
released in it.

How many particles stayed in the domain for ten years depended mainly
on the release depth (Table~\ref{tab:bg-domain}). More than 96\,\% of
the particles released below 100~m stayed in every decade, compared
with 52--93\,\% of those released at 0--50~m. Particles were lost when
they left the domain through its southern boundary or were removed at
the coastline. The remaining fraction therefore reflects both transport
and the treatment of boundaries.

\begin{table}[t]
\caption{Fraction (\%) of each release cohort still active north of
$50^\circ$N ten years after release, by decade of release and release
depth. Releases from 2010 onwards had mostly not reached ten years by
the end of the calculation and are omitted.}
\label{tab:bg-domain}
\centering
\begin{tabular}{lrrrr}
\tophline
decade of release & 0--50 m & 50--100 m & 100--250 m & 250--500 m \\
\middlehline
1970s & 65.4 & 88.8 & 97.7 & 99.7 \\
1980s & 52.3 & 80.5 & 96.4 & 99.5 \\
1990s & 81.0 & 91.2 & 98.2 & 99.8 \\
2000s & 93.4 & 96.7 & 99.0 & 99.7 \\
\bottomhline
\end{tabular}
\end{table}

\begin{figure*}[t]
\centering
\includegraphics[width=\textwidth]{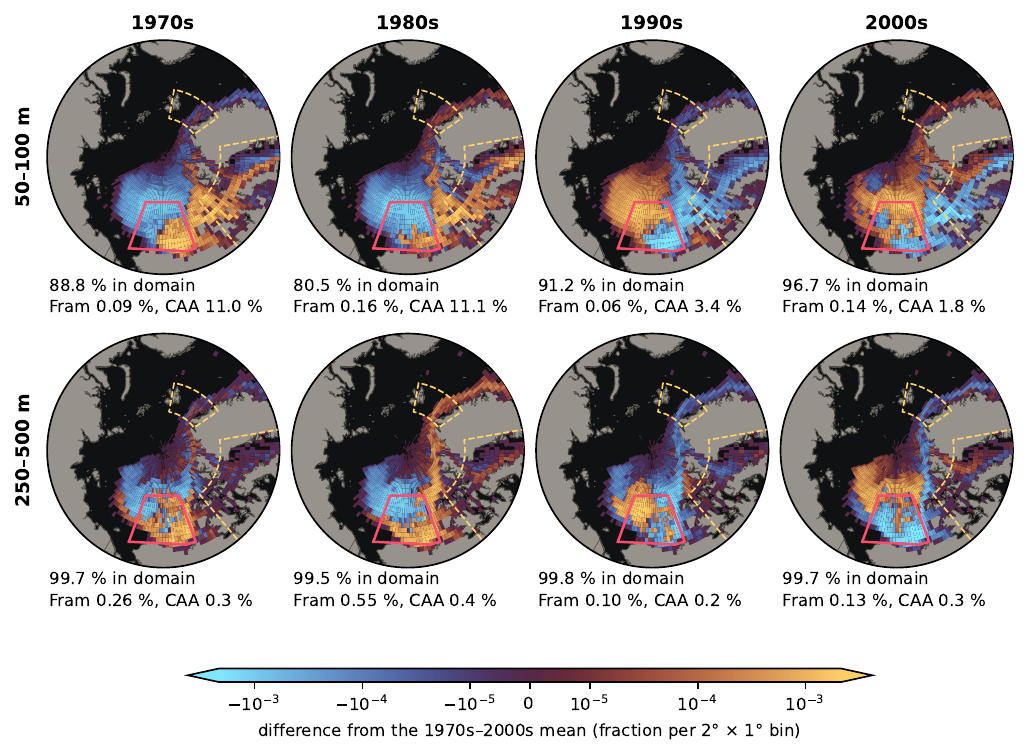}
\caption{Particle distribution ten years after release: difference
between the cohort of one decade (columns) and the mean of the four
decades, for releases at 50--100~m and 250--500~m (rows). Values are
fractions of the cohort per $2^\circ \times 1^\circ$ bin. The 1980s
column, for example, shows positions in 1990--1999. Orange bins hold
more particles than the mean, blue bins fewer. The colour scale is
logarithmic in both directions and linear between $-10^{-5}$ and
$10^{-5}$. Bins empty in the decade and in the
mean are not coloured. Below each panel: the fraction still active
north of $50^\circ$N and the fractions inside the dashed boxes (Fram:
Fram Strait, CAA: Canadian Arctic Archipelago and Baffin Bay). The solid
box marks the release region. Polar stereographic projection
north of $66^\circ$N.}
\label{fig:bg-presence}
\end{figure*}

Within the domain, the particle distribution after ten years differed
between release decades (Fig.~\ref{fig:bg-presence}). We binned the
ten-year positions on a $2^\circ \times 1^\circ$ longitude--latitude
grid and subtracted the 1970--2009 mean fraction in each bin for that
release depth.

For releases at 50--100~m, the later cohorts had fewer particles along
the route
through the Canadian Arctic Archipelago and Baffin Bay
(Fig.~\ref{fig:bg-presence}). After ten years, the box over this route
contained about 11\,\% of the 1970s and 1980s cohorts, but less
than 4\,\% of the 1990s and 2000s cohorts. The later cohorts
instead had more particles than the four-decade mean in the western
Canada Basin and towards the Chukchi Sea. Along northern Greenland
towards Fram Strait, the 1980s and 2000s cohorts had more particles
than the mean.

For releases at 250--500~m, the distributions also differed between decades,
although almost all particles remained in the domain
(Fig.~\ref{fig:bg-presence}). The 1970s cohort stayed closest to the
release region, whereas the 1980s cohort had the largest fraction in
the Fram Strait box. The 1990s and 2000s cohorts spread further west
and north of the release region.

\begin{figure*}[t]
\centering
\includegraphics[width=\textwidth]{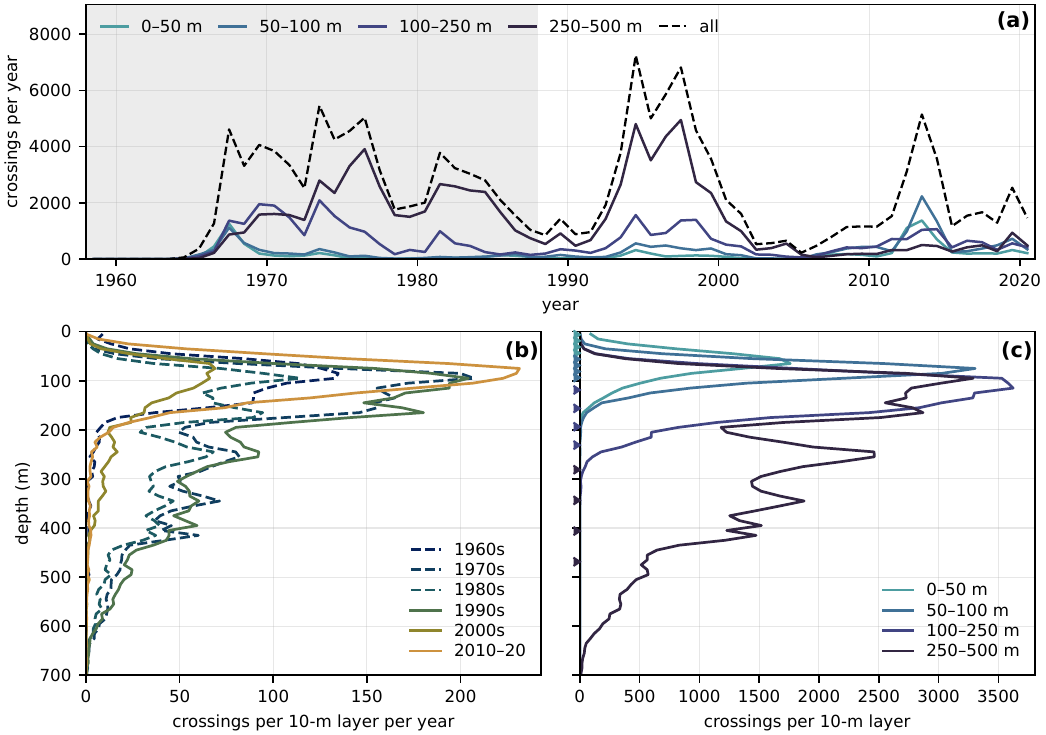}
\caption{Southward crossings of $79^\circ$N in Fram Strait
($25^\circ$W--$13^\circ$E), one per particle, without particles older
than 30 years at the crossing (9\,\% of all crossings). (a) Crossings
per calendar year by release depth, dashed: all release depths. Grey:
years before 1988, when not all ages up to 30 years were present.
(b) Crossings per 10-m layer and year, averaged over each decade of
the crossing (1960--1969, \ldots, 2010--2020), dashed: decades that
begin before 1988. (c) Crossings per 10-m layer over the whole
calculation, by release depth, triangles: the 16 release depths.}
\label{fig:bg-fram}
\end{figure*}

Positions ten years after release show where the particles are, but
not when they leave the Arctic. We therefore counted the first southward
crossing of Fram Strait for each particle (Fig.~\ref{fig:bg-fram}). The
section follows $79^\circ$N between northeast Greenland and Spitsbergen.
A crossing was counted when a particle moved southward across it between
two consecutive outputs. Of the 7.92 million particles, 2.0\,\% crossed
the section.

The particles reached the strait years after their release
(Table~\ref{tab:bg-fram-age}). For the releases of 1958--1980, each
followed for 40 years, the median age at the crossing was
14.4 years. Of the particles that crossed, 90\,\% did so within
29 years. The median age increased with release depth, from 7.9 years
at 0--50~m to 16.5 years at 250--500~m.

\begin{table}[t]
\caption{Age at the southward crossing of $79^\circ$N in Fram
Strait, by release depth, for the releases of 1958--1980, each followed
for 40 years. Crossed: percentage of the released particles. The age
columns refer to the particles that crossed. For the releases of
1958--1965, followed for 55 years, 2--8\,\% of the crossings occurred
more than 40 years after release.}
\label{tab:bg-fram-age}
\centering
\begin{tabular}{lrrrrr}
\tophline
release depth & released & crossed (\%) & median age (yr) & 90th percentile (yr) & older than 30~yr (\%) \\
\middlehline
0--50 m & 783\,748 & 0.55 & 7.9 & 13.3 & 5.1 \\
50--100 m & 725\,972 & 0.90 & 8.9 & 33.4 & 12.8 \\
100--250 m & 705\,272 & 3.38 & 11.5 & 21.0 & 4.4 \\
250--500 m & 675\,372 & 9.78 & 16.5 & 29.8 & 9.8 \\
all & 2\,890\,364 & 3.48 & 14.4 & 28.6 & 8.5 \\
\bottomhline
\end{tabular}
\end{table}

Because releases continued every three months throughout the
calculation, particles older than 30 years made a growing share of the
crossings, 13\,\% in the 1990s and 21\,\% in 2010--2020
(Table~\ref{tab:bg-fram-matrix}). To compare yearly crossings without
this growing share, Fig.~\ref{fig:bg-fram} includes only particles up
to 30 years old at the crossing. Once the calculation had run for
30 years, in 1988, each age up to 30 years was represented by the same
number of released particles, so the yearly counts from 1988 onwards
are directly comparable.

\begin{table*}[t]
\caption{Release decade of the particles that crossed $79^\circ$N in
Fram Strait southward, by decade of the crossing. Values are
percentages of the crossings of each decade (columns sum to 100), all
ages included. The last two rows give the number of crossings and the
percentage made by particles older than 30 years.}
\label{tab:bg-fram-matrix}
\centering
\begin{tabular}{lrrrrrr}
\tophline
release decade & 1960s & 1970s & 1980s & 1990s & 2000s & 2010--2020 \\
\middlehline
1958--1969 & 100.0 & 99.6 & 61.0 & 19.9 & 11.9 & 7.3 \\
1970s & -- & 0.4 & 38.5 & 29.5 & 14.6 & 8.2 \\
1980s & -- & -- & 0.5 & 49.0 & 27.2 & 10.5 \\
1990s & -- & -- & -- & 1.6 & 26.7 & 21.2 \\
2000s & -- & -- & -- & -- & 19.5 & 46.8 \\
2010--2020 & -- & -- & -- & -- & -- & 6.0 \\
\middlehline
crossings & 13\,685 & 35\,802 & 21\,943 & 46\,779 & 10\,901 & 30\,761 \\
older than 30~yr (\%) & 0.0 & 0.0 & 0.5 & 13.2 & 15.1 & 21.4 \\
\bottomhline
\end{tabular}
\end{table*}

From 1988 to 2020, crossings peaked in 1994--1997 and in 2013,
and were fewest in 2002--2005 (Fig.~\ref{fig:bg-fram}a). The number of
particles of these ages no longer changed, so these variations most
likely reflect the circulation. The crossings of the 1990s were deeper
than those of 2010--2020: 44 and 4\,\% occurred below 200~m
(Fig.~\ref{fig:bg-fram}b). Particles released at 0--50~m crossed a
median 37~m deeper than their release depth. Those released at
250--500~m crossed a median 120~m shallower (Fig.~\ref{fig:bg-fram}c).

The crossings of one decade came from releases of several decades
(Table~\ref{tab:bg-fram-matrix}). In the 1990s, particles released in
the 1980s, the 1970s and 1958--1969 made 49, 29 and 20\,\% of the
crossings.
Transport from the gyre to Fram Strait therefore took years to decades.
A change in the circulation of the gyre reaches the strait only
gradually, mixed with water released earlier and at other depths. This
agrees with \citet{Wang2024}: their tracer took several years to reach
Fram Strait, and a large part remained in the Arctic. Lagrangian
tracking by \citet{Lique2010} gave transit times of about a decade from
the gyre to the North Atlantic. The particle records resolve these
delays by release depth and decade.

\begin{figure*}[t]
\centering
\includegraphics[width=\textwidth]{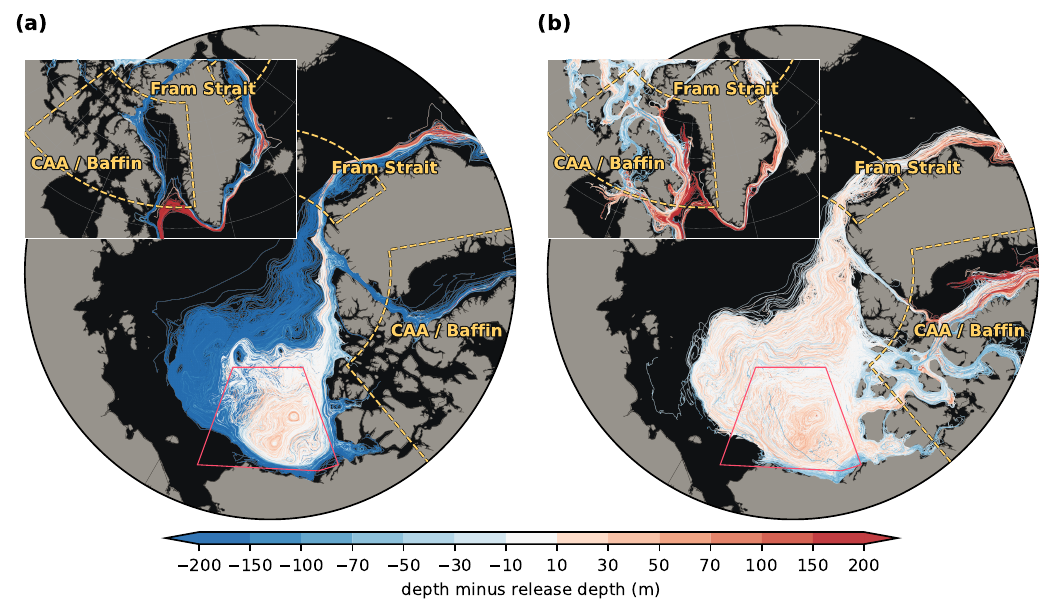}
\caption{Ten-year trajectories of particles released in the 1980s at
(a) 250--500~m and (b) 50--100~m depth. Every tenth particle of the 40
releases is drawn, from release to ten years later. Colour gives the depth relative to the release depth
(blue: shallower, red: deeper, range $\pm 200$~m). Insets enlarge the
region from the Canadian Arctic Archipelago to east Greenland. The red
box marks the release region, and dashed boxes mark the Fram Strait gateway
and the Canadian Arctic Archipelago and Baffin Bay gateway, which here
extends south to 63$^\circ$N (66$^\circ$N in
Fig.~\ref{fig:bg-presence}). Polar stereographic projections.}
\label{fig:bg-tracks}
\end{figure*}

Vertical motion along these routes differed between release depths
(Fig.~\ref{fig:bg-tracks}). Among the 1980s releases at 250--500~m,
most particles stayed in the Canada Basin. Those reaching the gateway
boxes were typically shallower than at release. Particles released at
50--100~m spread over a wider area and more often moved deeper.
Both depth groups included particles on the routes through the
Canadian Arctic Archipelago and along northern Greenland towards Fram
Strait (insets in Fig.~\ref{fig:bg-tracks}).

\subsection{Throughput and scaling}\label{sec:demo-performance}

This section examines how the cost of a run depends on particle count,
mesh size and the number of processes and nodes. Following
Eq.~(\ref{eq:cost}), each time step reads velocity fields, advances the
particles and writes trajectories. We first measured complete runs,
which include all three parts. To isolate the computational cost of tracking, we then measured it without field reads or trajectory output. Throughput
is reported in particle-steps per second. Rates for complete runs count
only active particles, those still being tracked.

All tests ran on DKRZ Levante nodes with two AMD EPYC 7763 processors
(128 cores) and 256~GB of memory, unless stated otherwise. The global
tests used ocean output from the nextGEMS IFS-FESOM production simulation
on the NG5 mesh, distributed on HEALPix grids
\citep{Rackow2025, Segura2025}.
The archive provides zooms 7 and 9. At zoom 9, the pixel spacing is
12.7~km, with 2.2 million wet surface nodes and 69 layers. For the
mesh-size tests, we resampled these fields to zooms 5--11, from 9418 to
35.5 million wet surface nodes. The resampling introduces no additional flow
structures, so these tests measure computational performance only. Particles were released at
every wet pixel centre at 100, 500 and 1500~m, which gave 6.0 million
particles at zoom 9. Tracking used RK2, linear temporal reconstruction
and a six-hour step, so each daily snapshot was used for four steps. Tracking
times are medians over steps without snapshot reads, taken for the
slowest process.

Tracking this zoom-9 release for one year took 2~h 12~min on one node,
including input and output. The run completed 1460 steps and wrote particle
positions once per day. Of the mean step time of 5.33~s, reading
velocity fields took 2.93~s, writing trajectories 1.78~s and tracking
0.62~s. Input and output thus accounted for 88\,\% of the step time. With
input and output included, the run advanced 0.84 million active particle-steps per second.
Tracking alone advanced 7.4 million. At the end of the year, 70\,\% of
the particles were still active. The trajectory output occupied 81~GB.

Because reading velocity fields dominates the cost, regional studies can
reduce this cost by limiting the volume of input data. KELP extracts the fields of
a region once, before tracking (Appendix~\ref{app:output}). In the
Beaufort Gyre experiment (Sect.~\ref{sec:demo-arctic}), the domain
north of $50^\circ$N contained 49\,154 of the 126\,858 nodes of the global
CORE2 mesh. The complete 63-year run with 252 releases took 3~h 36~min
on one node, including input and output. It advanced 5.0 million
active particle-steps per second, with a one-day step and 32 processes
with four threads each. This rate is not directly comparable with the global
run, because the mesh, release and time step differ.

\begin{figure*}[p]
\centering
\includegraphics[width=\textwidth]{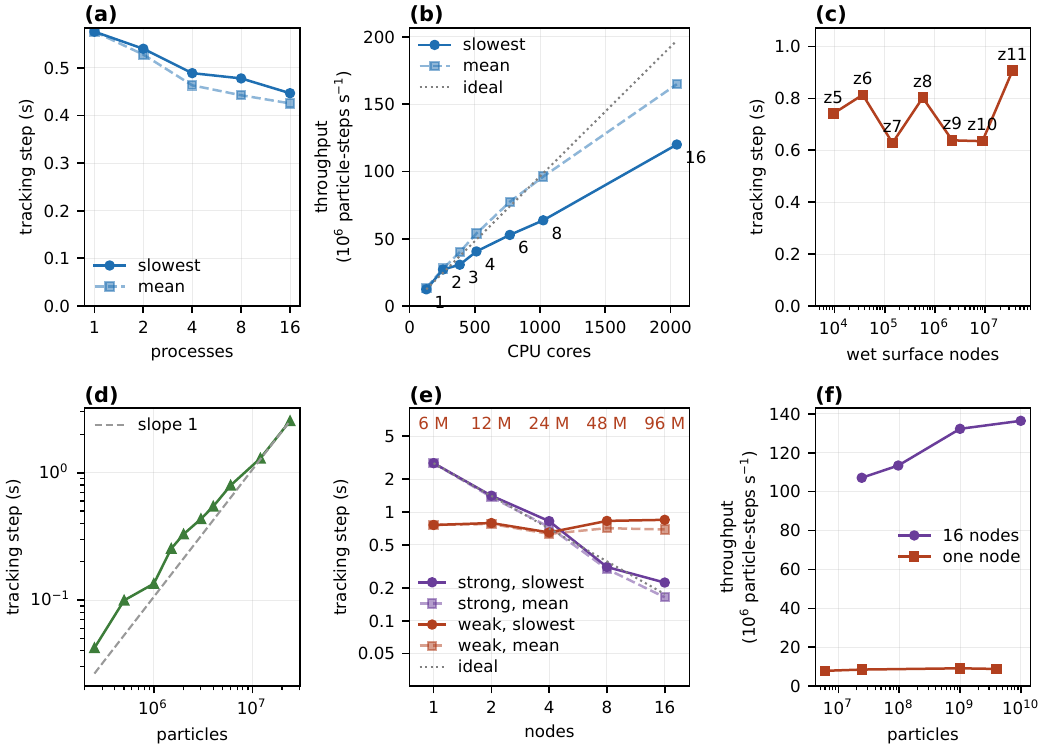}
\caption{Scaling on ng5 HEALPix fields. Fields at zooms 7 and 9 came
from the archive. The other zooms were resampled from them and contain
no additional flow structures. (a) Process layouts for 6.0
million particles at zoom 9 on one 128-core node, threads per process
equal 128 divided by the number of processes. Circles show the slowest
process and squares the mean over processes. (b) Throughput for the
same particles on 1--16 nodes with eight 16-thread processes per node,
labels give the number of nodes, symbols follow panel (a), and the
dotted line shows ideal scaling from one node. (c) Tracking time across
zooms 5--11 (labels) for the same particles and one 128-thread process.
(d) Tracking time versus particle count at zoom 10, one 128-thread
process, and the guide has slope 1. (e) Zoom 11 with one 128-thread process
per node: strong scaling of 24 million particles and weak scaling with
6.0 million particles per node (totals along the top). (f) Throughput
at zoom 11 versus the number of particles tracked at once, on sixteen
nodes and on one node with 1~TB of memory.}
\label{fig:e2-scaling}
\end{figure*}

The cost of tracking alone depended much more on particle count than on
mesh size (Fig.~\ref{fig:e2-scaling}c, d). As the mesh size increased
3765-fold from zoom 5 to zoom 11 for the same 6.0 million particles,
the tracking time per step varied by at most a factor of 1.45, which
corresponds to 6.6--9.6 million particle-steps per second on 128
threads. The variation between zooms was not monotonic and has no
identified cause. At zoom 10, tracking time grew nearly in proportion to
particle count from 0.25 to 24 million particles (log--log slope 0.88).

Distributing the release among processes increased throughput within and
across nodes (Fig.~\ref{fig:e2-scaling}a, b). On one node, sixteen
eight-thread processes achieved 29\,\% higher throughput than one
128-thread process. With eight 16-thread processes per node, throughput
rose from 12.3 million particle-steps per second on one node to
120 million on sixteen nodes. The parallel efficiency was 61\,\% for
the slowest process and 76\,\% for the mean over processes. On sixteen
nodes, each process followed only 47\,000 particles, and the slowest
particle group determined the completion time.

The zoom-11 grid, with a pixel spacing of 3.2~km and 35.5 million wet
surface nodes, tested scaling on a large mesh
(Fig.~\ref{fig:e2-scaling}e). One process required about 160~GB of memory
for fields, mesh and read buffers, so only one process fitted on a
node. With 6.0 million particles per node, sixteen nodes tracked sixteen
times as many particles in 1.12 times the time of one node, a
weak-scaling efficiency of 90\,\%. The resulting 96.2 million
particles reached 113 million particle-steps per second. For 24
million particles, strong scaling gave a 12.5-fold speed-up on sixteen
nodes. At this zoom, reading one daily snapshot took about 146~s per
process. This is equivalent to about 190 tracking steps for 6.0 million particles.

Short tests showed how many particles a run can hold
(Fig.~\ref{fig:e2-scaling}f). On sixteen nodes, $10^9$ and $10^{10}$
particles reached 132 and 136 million particle-steps per second. The
larger test used nodes with 512~GB of memory. One node with 1~TB of
memory tracked $4 \times 10^9$ particles at 8.9 million particle-steps
per second. In addition to a fixed part of about 66~GB per process, memory use grew
by about 180 bytes per particle. These tests ran three or four
steps without trajectory output. They therefore measure tracking
capacity, not the cost of a complete study.

\subsection{Setting up studies with an agent}\label{sec:demo-agent}

A coding agent prepared and ran all studies of this paper with the KELP
skills (Table~\ref{tab:skills}) and wrote the scripts that processed
their output. The agent carries out each preparation step in a short
turn and then pauses until the user answers open questions on the
setup, checks intermediate output or approves a cluster launch. The
times given below are therefore calendar times from the initial request
to the completed setup or run, paced by these user checkpoints, data
extraction and batch queues rather than by continuous agent work.

For the mesh test of
Sect.~\ref{sec:ver-mesh}, the agent built the nine test meshes, the
exact flow and the run scripts, and the results were available about
2~h after the request and the figures and tables the same evening. For the comparison of model-mesh and HEALPix fields
(Sect.~\ref{sec:demo-healpix}), the agent prepared 16 configurations and the
release files in one morning. The year-long runs, their processing and
the figures were complete the same day.

New data and a second tracker took similar time. For the wind-farm
comparison (Sect.~\ref{sec:ver-reference}), the FESOM-C output had been
stored in a non-standard format. The agent inspected 1.4~TB of it and
wrote and tested an input module for it on the first day. The
production runs and the first maps followed the next day. For the
comparison with Parcels, it set up that code with matched
configurations in about 3~h.

For the Beaufort Gyre
(Sect.~\ref{sec:demo-arctic}), preparing the releases and a restartable
chain of jobs spanned two calendar days, including the extraction of
63 years of regional fields, restart checks and user approvals between
steps. For the scaling tests (Sect.~\ref{sec:demo-performance}), the agent
generated and checked 370 configurations in about 1~h, and the 14 jobs
ran overnight. The agent wrote a report for each campaign with its inputs,
configurations, runs and results.

The agent also found problems in the engine itself. On the wind-farm
mesh, a run was 100 times slower than expected. In three-dimensional
HEALPix tracking, one code path was slow, and in the scaling tests a
defect limited the mesh size. In each case, the agent traced the
problem to its cause in the code. The fixes entered the engine before
the runs reported here. None of these problems produced an error
message. They appeared as slow runs or unexpected limits, which are
easy to overlook in routine work.

An agent can also prepare a tracking study from the code and its
README alone. To measure what the skills add, we gave agents the same tasks
with and without them and graded the files they produced. Seven tasks
were set, each described in a short request as a user would write it. They included surface and backward tracking, coastal stranding and the
extraction of a region with restarts. Two further tasks were the
analysis of a trajectory archive and the diagnosis of a slow
configuration. The most
complex task asked for a multiyear campaign around Greenland, with
quarterly releases at several depths and ten-year trajectories. The
agents prepared the files, could examine the data and could check their
configurations with KELP, but no tracking calculation was started.

Each task was given under two conditions, which differ in the files
the agent could read. Under the README condition, the agent had the code with its README,
examples, utility scripts and tests. Under the Full condition, it also had the topic documentation, the
reference manual and the seven skills, with the short guide files that
point an agent to them. For the Claude models, a third condition, Docs, added only the
documentation and the manual to README. Every attempt started from a
fresh copy of the same code revision, and the request and the expected
files were the same under all conditions.

Two command-line coding agents were tested, Claude Code (Anthropic,
version 2.1.295) and Codex (OpenAI, version 0.161.0). Claude Code ran
with three Anthropic models, Haiku 5.5, Sonnet 5.5 and Opus 5.5. Codex
ran with two OpenAI models, gpt-6-luna and gpt-6-sol. Each of these
five models made two attempts at each task under each condition. This
amounted to 140 attempts under README and Full together and 42 Claude
attempts under Docs.

The files of each attempt were graded by a separate language model,
Fable 5.1 (Anthropic), which was not among the five models under test. It received the
files without names and without the record of the attempt, and it
checked them against 98 criteria written for the seven tasks. Eleven of the criteria concern the working rules of the project. Two
examples are checking the sign convention of the vertical velocity in
the input data and keeping a record of each job. A criterion was classed as critical
when failing it would prevent a run or give a wrong result. The
preparation score of a task is the percentage of criteria met, averaged
over the attempts of one model under one condition
(Table~\ref{tab:agent-bench}). The overall score gives equal weight to
each of the seven tasks.

All models ran at medium reasoning effort, the Codex default. We chose
this level after a small pilot study with the Claude models on two of
the tasks. Under Full, medium and high effort gave nearly identical
scores, but high effort cost up to 2.7 times as much. Low effort
reduced the score of Haiku. Without Full, the scores varied more with
effort.

\begin{table*}[t]
\caption{Preparation score by task, coding agent, language model and
documentation condition, and mean token use per attempt. README: code,
examples, utilities and tests. Full: adds topic documentation, the
manual and the skills. The score is the mean
percentage of grading criteria met across attempts for each task and
condition. Overall is the equally weighted mean of the seven
task scores. Input tokens include repeated cached context (more than 90\,\%). The Claude and OpenAI models count tokens with
different tokenizers. All models ran at medium reasoning effort.}
\label{tab:agent-bench}
\centering
\footnotesize
\setlength{\tabcolsep}{3pt}
\begin{tabular}{>{\raggedright\arraybackslash}p{4.4cm}rrrrrrrrrr}
\tophline
& \multicolumn{6}{c}{Claude Code} & \multicolumn{4}{c}{Codex} \\
\cline{2-7}\cline{8-11}
& \multicolumn{2}{c}{Haiku 5.5} & \multicolumn{2}{c}{Sonnet 5.5} & \multicolumn{2}{c}{Opus 5.5}
& \multicolumn{2}{c}{gpt-6-luna} & \multicolumn{2}{c}{gpt-6-sol} \\
& README & Full & README & Full & README & Full
& README & Full & README & Full \\
\middlehline
\multicolumn{11}{l}{Score (\%)} \\
Surface tracking and presence map & 89 & 100 & 93 & 96 & 100 & 100 & 89 & 96 & 93 & 93 \\
3-D backward tracking & 93 & 100 & 93 & 100 & 93 & 100 & 87 & 90 & 93 & 100 \\
Coastal stranding & 89 & 100 & 89 & 93 & 96 & 100 & 82 & 82 & 100 & 100 \\
Regional extraction and restarts & 50 & 97 & 76 & 100 & 85 & 100 & 41 & 97 & 65 & 100 \\
Fram Strait crossings from archive & 95 & 86 & 91 & 100 & 91 & 100 & 73 & 100 & 100 & 100 \\
Slow-configuration diagnosis & 100 & 100 & 95 & 90 & 100 & 100 & 95 & 85 & 95 & 100 \\
Multiyear Greenland campaign & 71 & 94 & 85 & 100 & 85 & 100 & 56 & 65 & 71 & 100 \\
\middlehline
Overall & 84 & 97 & 89 & 97 & 93 & 100 & 75 & 88 & 88 & 99 \\
\middlehline
\multicolumn{11}{l}{Mean token use per attempt} \\
Input ($10^6$ tokens) & 3.5 & 3.7 & 2.8 & 2.3 & 4.1 & 4.8 & 0.85 & 0.99 & 1.3 & 1.9 \\
Output ($10^3$ tokens) & 63 & 53 & 36 & 25 & 41 & 40 & 9 & 11 & 12 & 14 \\
\bottomhline
\end{tabular}
\end{table*}

Compared with README, Full raised the mean preparation scores of all
five language models by 7 to 13 percentage points
(Table~\ref{tab:agent-bench}). Each mean is based on seven tasks and 14
attempts.
The improvement varied by task. It was largest for regional extraction
with restarts and for the multiyear Greenland campaign. A few task
scores decreased with Full. Each task score is a mean of two attempts, so a single unmet criterion
can shift it by several percentage points.

Adding documentation alone, in the Docs condition tested with the
Claude models, changed the mean scores by at most one percentage point
compared with README. The improvement with Full therefore came mainly
from the skills, and Table~\ref{tab:agent-bench} compares README and
Full directly. Each model read KELP skills in most Full attempts. The
largest gain concerned the working rules of the project, which the
agents often missed under README but followed almost completely under
Full, and all five models also improved on the remaining criteria.

Full reduced critical errors but did not eliminate them. With README, 54 of the 70 attempts failed at least one critical
criterion. With Full, this number decreased to 18. Agent-prepared
studies need to be checked by the user, with or
without skills.
Overall, adding the skills improved the
preparation of KELP studies by all five language models.

\section{Discussion}\label{sec:discussion}

One tracking engine served all experiments of this paper, from flow
around a single wind-turbine foundation to multidecadal Arctic transport
and global releases. No separate tracker was needed for FESOM output on
its own mesh or for HEALPix fields. The wind-farm verification and the
Beaufort application use triangular model meshes, while the HEALPix experiments
use a dual mesh. The analytic and mesh-quality tests cover every supported mesh type,
including the quadrilateral and mixed meshes not used in the ocean-model
applications here (Sects.~\ref{sec:ver-analytic} and
\ref{sec:ver-mesh}).

The choice between the model mesh and HEALPix fields depends on the
transport quantity of interest. In the comparison on CORE2, absolute
dispersion agreed with the model-mesh value more closely than
individual trajectories did (Sect.~\ref{sec:demo-healpix}). Agreement in
a global statistic therefore does not establish local pathway accuracy,
especially near coasts and strong currents. The same engine lets us
compare both representations with releases and numerical settings held
fixed. The measured differences then combine remapping, spatial
interpolation and coastline representation.

Time-step refinement must likewise be assessed through the chosen
transport diagnostic. The temporal tests show why smaller integration
steps can leave errors from velocity reconstruction unchanged
(Sect.~\ref{sec:ver-order}). For atmospheric trajectories, temporal
interpolation of the wind caused larger errors than spatial
interpolation under typical conditions \citep{Stohl1995}. In an
eddy-resolving ocean model, offline particles spread less vertically
than an online tracer, partly because subdaily vertical velocities were
not resolved \citep{Wagner2019}. Near structures resolved by the mesh,
the time step also changes how particles pass through the local flow.

\begin{figure*}[t]
\centering
\includegraphics[width=\textwidth]{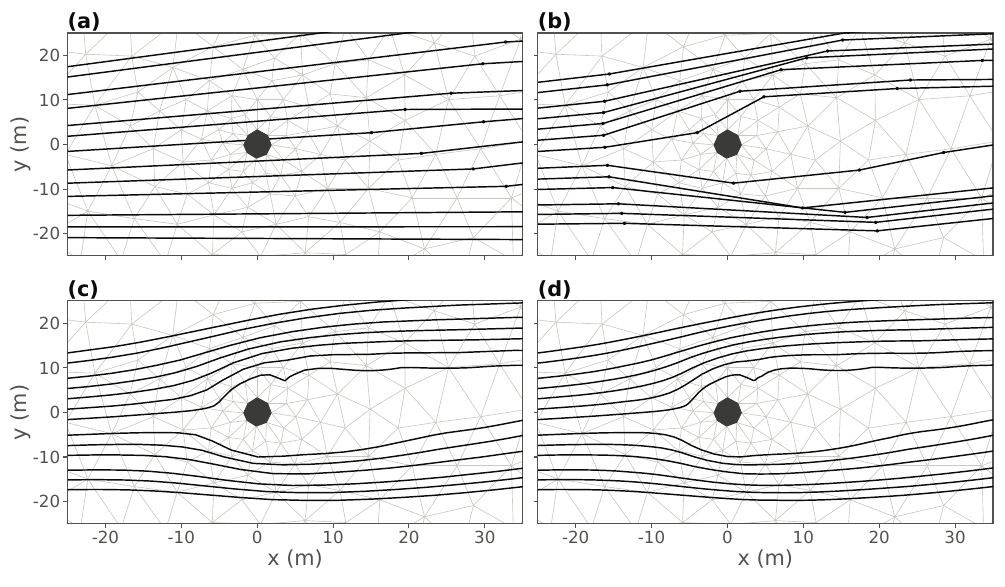}
\caption{Particle paths around a wind-turbine foundation, computed with
RK4 and PCHIP at integration steps $h$ of (a) 300, (b) 75, (c) 9.375 and (d) 4.69~s. The same 13
particles are shown during the first 2~h after release. Grey lines show
mesh edges and the dark polygon the foundation. Dots mark integration
steps in (a) and (b), and lines join successive positions. Axes give
distances from the foundation centre in metres.}
\label{fig:wf-passby}
\end{figure*}

The wind-farm setup of Sect.~\ref{sec:ver-reference} illustrates this
local sensitivity. We released a north--south line of 301 particles
300~m west of one foundation and tracked it at steps from 300 to
4.69~s. In the eastward flood flow (about
0.4~m\,s$^{-1}$), a 300-s step spans about 120~m and can cross the
foundation and its wake (Fig.~\ref{fig:wf-passby}a). At finer steps, the
paths curve around the foundation. The paths at 9.375 and 4.69~s nearly
coincide (Fig.~\ref{fig:wf-passby}c, d).

These local path differences also affected transport diagnostics over
longer intervals. We repeated all releases of the wind-farm run with RK4
and PCHIP at six steps, halved from 300 to 9.375~s. Each release was
followed for 29.5 days. The first diagnostic was the fraction of
positions that moved less than 10~m over the next 2~h. The second was
the area holding 90\,\% of the probability.

In the three clouds inside wind farms, both diagnostics changed
non-monotonically with the integration step. Only the two finest steps,
18.75 and 9.375~s, agreed in all three clouds. Without wind farms, the
areas of these clouds were the same at steps from 300 to 37.5~s.

For multidecadal releases, transport comparisons also depend on the age
of the particles. Comparing Beaufort cohorts at the same age separates
their release history from the duration of drift
(Sect.~\ref{sec:demo-arctic}). Likewise, yearly crossing counts are
comparable only within a fixed range of particle ages, because older
particles join the population throughout the calculation. Particle
disappearance, however, is not by itself evidence of export: particles may leave the extracted domain or be lost by the cell search. Recorded crossings
measure passage through a chosen section, while the trajectories retain
the routes and vertical motion leading to it.

The cost of a complete study depends on data handling as well as
tracking throughput. Short tracking tests reached billions of particles,
but input and output dominated the full-year experiment
(Sect.~\ref{sec:demo-performance}). A timing sample therefore needs the
intended fields, release and output interval. The number of particles is
not limited by the engine. An ensemble is divided among processes on one
or several nodes. Each process needs a fixed amount of memory for the
fields and about 180 bytes per particle
(Sect.~\ref{sec:demo-performance}). The $10^{10}$ particles of the
largest test therefore demonstrate the capacity of sixteen nodes, not a
limit of the code. The practical limit is the memory and the output
volume that a study can afford.

Large ensembles are limited by the writing and storage of their
trajectories. The present applications compute their diagnostics from
saved trajectories. Presence counts, section crossings and
connectivity estimates could instead be computed during the run, so
that only selected trajectories need to be written. Connectivity
estimates are used, for example, in studies of larval dispersal
\citep{CowenSponaugle2009}. They depend on the number of released
particles and on the size of the regions compared
\citep{Jones2016, Monroy2017}. Complete archives remain useful when
several questions are addressed afterwards, as in the Beaufort
analysis.

The skills direct agents to keep the records needed to repeat a study
independently of the conversation that produced it. These records
include releases, configurations, commands and analysis scripts,
together with the input data and code revision used
(Sect.~\ref{agent-layer}). They also retain the setup and analysis
decisions when one experiment grows into a series. The wind-farm
time-step series and the scaling campaigns are examples.

With skills, the wind-farm time-step series required little of the user's time. An agent prepared the configurations, batch scripts
and comparison scripts. The user set the task, decided the setup
questions and approved each launch (Sect.~\ref{agent-layer}). A
convergence test of this kind can therefore precede the physical
interpretation of a transport diagnostic.

The controlled evaluation connects the skills with the completeness of
prepared study files (Sect.~\ref{sec:demo-agent}). All five language
models prepared more complete files with skills, most clearly for
regional extraction and the multiyear Greenland campaign. Their improvements extended beyond the criteria for the working rules of
the project.
For the three Claude models, documentation alone changed the mean scores
little. A likely reason is that the documentation describes how the
engine works, whereas the skills state the steps and checks of a task.
Some attempts with skills still contained errors that would stop a run
or give a wrong result. The user's checks therefore remain part of
an agent-prepared study.

KELP's scope complements existing approaches to particle tracking.
Parcels v4 also provides access to FESOM2 output
\citep{ParcelsTeam2026a}. On such output, the two codes gave closely
agreeing trajectories with matched settings
(Sect.~\ref{sec:ver-reference}). KELP combines that access with
quadrilateral and mixed coastal meshes, HEALPix grids, large-ensemble
execution and agent skills. Other
frameworks provide functions outside its present implementation:
TRACMASS assigns transport to trajectories \citep{Doos1995, Doos2017}.
OpenDrift includes oil-weathering and pelagic-egg modules
\citep{Dagestad2018}, and Parcels accepts particle behaviour as Python
kernels \citep{LangeVanSebille2017}. OceanTracker computes heat maps and
connectivity during the run, resuspends particles from the seabed and
nests regional meshes \citep{Vennell2026}.

On unstructured meshes, OceanTracker targets coastal models such as
SCHISM, FVCOM and Delft3D-FM, splits quadrilateral cells into two
triangles and interpolates velocities from the vertices
\citep{Vennell2026}. KELP keeps quadrilateral cells whole and supports
vertex velocities, with Wachspress interpolation on general
quadrilaterals, and cell-centred velocities, as in FESOM2 and HEALPix
output. It covers coastal domains and global domains that span the
dateline and the poles (Sect.~\ref{sec:ver-analytic}).

OceanTracker runs threads in the shared memory of a single machine
\citep{Vennell2026}. KELP likewise runs multithreaded on one
workstation or node, as in the coastal and regional runs of
Sects.~\ref{sec:ver-reference} and \ref{sec:demo-arctic}. It also
distributes processes across cluster nodes when the ensemble size or
the fields of a global grid exceed the memory of one machine
(Sect.~\ref{sec:demo-performance}). On the finest global HEALPix grid
of that section, only one process fitted into the 256\,GB memory of a
node. These frameworks are operated through configuration files or
Python scripts, and KELP adds skills for coding agents and the records
that make a study repeatable (Sect.~\ref{sec:software}).

Several limits of the present version should be kept in mind. KELP has
no mechanism that steers particles around islands and coasts. A particle
that the interpolated velocity carries onto land is removed or, with
stranding enabled, kept on the shore. The boundary treatment of the
plane fit over edge-sharing neighbours reduces the velocity towards the
coast but does not deflect the particle (Appendix~\ref{app:interp}).
Stranding and drying are decided from the water depth at the nearest
node and from the coastal cells (Appendix~\ref{app:stranding}). They
are rules, not a model of the shore. Drying is taken as the model output
gives it, and in regional extracts without sea-surface height it is not
represented at all. The stranding rules were checked only in analytic
tests, not in the experiments of this paper. The default cell search,
two rings of cells around the nearest node, loses particles on strongly
stretched and distorted quadrilateral and mixed cells
(Sect.~\ref{sec:ver-mesh}). Such cells are rare in the model meshes we
measured, and the edge walk keeps the particles where they occur.

KELP computes every particle position in double precision and can keep
the velocity fields in single precision to halve their memory. Single
precision for the positions themselves would not suffice near coastal
structures. At $54^\circ$N, one unit in the last place of a
single-precision longitude is about 0.25~m. A particle in the wind-farm
flow moves about 1.9~m in the finest step of
Sect.~\ref{sec:ver-reference}. Rounding would then change each step by several per cent. In the
comparison of Sect.~\ref{sec:ver-reference}, the Parcels runs kept
single-precision positions, and their forward--backward closure was
1.6~m against 0.1~mm in KELP.

Comparison with observed drifters and validation of backward stochastic
runs as source-location distributions \citep{Thygesen2011} are outside
the present verification. The limits of each input module for diffusion, adaptive integration
and coordinate frames are given in Sect.~\ref{numerical-methods}, Table~\ref{tab:hosts} and
Appendix~\ref{app:frame}.

\conclusions\label{sec:conclusions}

We presented KELP v1.0, a Lagrangian particle tracker that combines
tracking on triangular, quadrilateral and mixed ocean-model meshes and
on HEALPix grids with parallel execution for large ensembles and
operation by coding agents. Its numerical schemes are established. To
them KELP adds configuration checks, saved records and seven task
skills, so that a coding agent can prepare, run and process a study
while the user sets the task, approves each launch and checks the
results. All experiments in this paper were prepared and run in this way, with
setups completed in hours to two calendar days including user reviews
and test runs. In the controlled evaluation, the skills raised the mean
preparation scores of five language models from 75--93\,\% to
88--100\,\% (Sect.~\ref{sec:demo-agent}). The errors that
remained show that the user's checks stay part of an agent-prepared
study.

Verification against two other trackers on the same fields gave a
95\,\% overlap of the particle distributions with the tracker of
\citet{Androsov2026} on a coastal wind-farm mesh and a median separation
of 58~m after one year from Parcels v4 on the cropped global CORE2 mesh
(Sect.~\ref{sec:ver-reference}). The mesh tests showed that three of the
four interpolation schemes keep second-order accuracy on distorted
cells, whereas averaging centroid velocities to the nodes gives 9--18
times larger errors there, and that the edge walk loses no particles on
stretched cells where the default cell search loses 24--61\,\%
(Sect.~\ref{sec:ver-mesh}). The time-step tests showed that cubic
temporal reconstruction reduces the error between stored snapshots
15-fold, which a higher-order integrator alone cannot do
(Sect.~\ref{sec:ver-order}).

Each halving of the HEALPix pixel spacing reduced the median separation
from the model-mesh trajectories by a factor of about three, to 1.6~km
after one year at 25.5~km, while absolute dispersion agreed within about
2\,\% already at 51~km. HEALPix output therefore serves studies of
ensemble transport statistics, and the model mesh remains the reference
for individual pathways (Sect.~\ref{sec:demo-healpix}). In the Beaufort
Gyre study, regional extraction and restart files let 7.92 million
particles from 252 quarterly releases be tracked over 63 years in 3~h
36~min on one 128-core node (Sects.~\ref{sec:demo-arctic} and
\ref{sec:demo-performance}). The trajectories showed that transport from
the gyre to Fram Strait takes years to decades: the median age at the
crossing was 14.4 years and rose from 7.9 years for releases at 0--50~m
to 16.5 years at 250--500~m, so that the export of one decade mixes
water released over several earlier decades, and release cohorts must
be compared at the same age. Scaling tests reached $10^{10}$ particles
on sixteen nodes, and in a full-year global run field input and
trajectory output took 88\,\% of the step time
(Sect.~\ref{sec:demo-performance}).

The limits of the present version are listed in
Sect.~\ref{sec:discussion}. Computing transport diagnostics during the
run is the main development priority, so that large ensembles need not
write complete trajectory archives. Further work will extend spatially
varying diffusion to three dimensions and broaden the scope of adaptive
integration.

\appendix
\section{Numerical details}\label{app:numerics}

\subsection{Locating particles}\label{app:localization}

The lookup grid of Sect.~\ref{one-time-step} assigns the mesh nodes to
square boxes in longitude and latitude. The initial box side is proportional to \(\sqrt{A/N}\), where \(A\)
is the area of the mesh bounding box in square degrees and \(N\) the
number of nodes. By default, KELP also tests finer boxes on a sample of
cell centres and adopts a smaller box when that lowers the estimated
search cost by a set factor, or it uses a box size given in the
configuration.

\begin{figure*}[t]
\centering
\includegraphics[width=\textwidth]{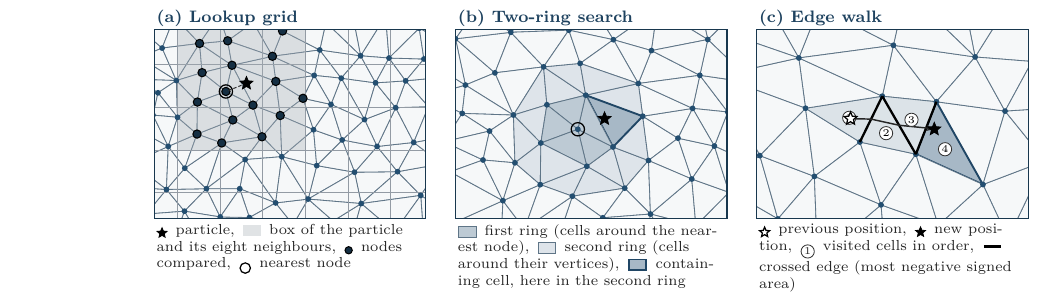}
\caption{Locating a particle on a triangular patch. (a) Nearest-node search: the
particle, the box of the particle with its eight neighbours, the nodes
compared and the nearest node. (b) Two-ring search: the first ring
around the nearest node, the second ring around vertices of first-ring
cells and the containing cell. (c) Edge walk from the previous position
to the new one: the visited cells in order and the crossed edges.}
\label{fig:localization}
\end{figure*}

For each particle position, the search first scans the box that
contains it and the eight neighbouring boxes
(Fig.~\ref{fig:localization}a). It compares chord distances on the unit
sphere. The distance to the
nearest candidate defines a spherical cap centred on the particle. If
the searched window
covers this cap, no unsearched node can be closer. Otherwise, the window
expands within a search limit derived from the longest mesh-cell edge.
The search accounts for the dateline and includes every longitude when
the cap reaches a pole. By default, the combined routine and the
stage-by-stage execution of Sect.~\ref{data-flow} share this search and
resolve ties in the same order.

The containing cell is sought in two rings around the nearest node
(Fig.~\ref{fig:localization}b). The second ring contains cells attached
to vertices of first-ring cells. For triangles, containment is checked
from the signs of the barycentric coordinates. Quadrilaterals are split
along a diagonal into two triangles that are tested in the same way,
with tolerance \(-10^{-6}\) in normalized coordinates. If
neither ring yields a containing cell, the position is classified as
outside the mesh.

The edge walk starts from the previously occupied cell and crosses the
edge with the most negative normalized signed area
(Fig.~\ref{fig:localization}c). The walk stops once all signs indicate
containment. The two-ring search takes over if the walk reaches a boundary
edge or a degenerate cell. It also takes over if the walk crosses the
same edge twice or exceeds 256 crossings. The two paths agree where the
two-ring search is exhaustive. On strongly distorted meshes, the walk can
reach a valid cell beyond the second ring.

\subsection{Horizontal interpolation}\label{app:interp}

Horizontal interpolation is performed in local planar coordinates. For
longitude and latitude expressed in degrees, the coordinates are:
\begin{equation}\label{eq:plane}
  x_k = \Delta\lambda_k \,\frac{\pi}{180}\, R \cos\phi_0, \qquad
  y_k = (\phi_k - \phi_0) \,\frac{\pi}{180}\, R ,
\end{equation}
where \(R\) is the Earth radius and \(\Delta\lambda_k\) is the longitude
difference wrapped to \([-180^\circ,180^\circ)\). Neither the weights
nor the fitted value depends on \(R\). The frame is centred at
\((\lambda_0,\phi_0)\): the cell centroid for the interpolation from the vertices, or the
particle position for the plane fit.

On triangles, the interpolation from the vertices is
\begin{equation}\label{eq:bary}
  (U, V) = \sum_{i} w_i\,(u_i, v_i), \qquad
  w_i = \frac{\mathrm{area}(\text{sub-triangle opposite vertex } i)}{\mathrm{area}(\text{triangle})},
\end{equation}
with nodal velocities \((u_i,v_i)\) and \(\sum_i w_i=1\). Convex
quadrilaterals use Wachspress weights \citep{Warren1996}.
Both methods reproduce fields that are stored at nodes and vary linearly
in the local coordinates. When velocities are stored at cell centroids,
vertex values are first calculated as area-weighted means over
neighbouring cells. This averaging preserves constant fields. Linear
fields are reproduced exactly only where the area-weighted centre of the
surrounding centroids coincides with the node.

The plane fit is used for velocities stored at the cell centroids. It
minimizes
\begin{equation}\label{eq:lsq}
  \min_{a_0, a_1, a_2} \sum_{j \in S} \bigl( a_0 + a_1 x_j + a_2 y_j - f_j \bigr)^2 ,
\end{equation}
for each velocity component \(f\), where \(f_j\) is the centroid value
of cell \(j\) at the local coordinates \((x_j,y_j)\). The fitted value
at the particle is \(a_0\). The set of cells \(S\) is either the cells
around the nearest node or the containing cell and its edge-sharing
neighbours. The fit is singular when the centroids lie on one line. The particle
is then returned to its position before the step and marked as
removed.

At the coast, the neighbour across a coastal edge is missing. Its place is taken by a mirror image of the containing cell across the
coastal edge. The velocity of the image is derived from that of the
containing cell.
With free slip, the component normal to the coast is reversed and the
tangential component kept, so that a flow along the coast is
reproduced. With no slip, both components are reversed.

\subsection{Vertical interpolation and clamping}\label{app:vertical}

Vertical interpolation first samples the column at each node, then
combines these values horizontally. Fixed-level tracking uses a common
interface table and the floor depth at each node. For sigma layers,
interface depths are
\begin{equation}\label{eq:sigma}
  H = \max(d + \eta,\, D_{\min}), \qquad z_k = (1 - \sigma_k)\, H ,
\end{equation}
where \(\sigma_k\) gives the fractional coordinate of each interface,
\(d\) is the floor depth and \(\eta\) is the sea-surface height,
positive upward. The minimum column height \(D_{\min}\) defaults to
0.5~m.

Vertical velocity is interpolated between interfaces as
\begin{equation}\label{eq:wlin}
  W(z) = (1 - \alpha)\, W_k + \alpha\, W_{k+1}, \qquad
  \alpha = \frac{z - z_k}{z_{k+1} - z_k} ,
\end{equation}
with no extrapolation outside the local column. Horizontal interpolation
of \(W\) uses a plane fit, the nearest-node value, the interpolation from the
vertices or
inverse-distance weighting \citep{Shepard1968}. On sigma meshes, each
neighbouring column is sampled at the particle's fractional depth. The
local column height \(H_p\) is calculated with the same horizontal
interpolation as \(W\). The resulting sampling depth at node \(j\) is
\(z_j=(z/H_p)H_j\).

In global FESOM output, the layer interfaces move with the sea-surface
height \(\eta\). The vertical velocity \(W\) written by the model is
the flow through these moving interfaces. It is not the rate of
change of the depth below a fixed datum. KELP advances the particle in
the reference depth, the depth the interfaces have when \(\eta=0\),
and uses \(W\) directly as its rate of change. The exact rate differs from \(W\) by the factor \(H/(H+\eta)\), where
\(H\) is the water depth. The error in the vertical rate is therefore
of relative order \(\eta/H\). The reference depth itself differs from the physical depth
by at most \(|\eta|\).

On sigma meshes, the same idea is applied to the fraction of the water
column. KELP advances the fractional depth \(s=z/H_p\) through
\(\mathrm{d}s/\mathrm{d}t=W/H_p\), where \(W\) is the flow through the
sigma surfaces that the model writes. Each Runge--Kutta stage evaluates
the column height at that stage's position and time and then obtains
the depth as \(z=sH_p\). When the flow through the sigma surfaces is
zero, \(s\) stays constant. The particle then rises and sinks with the
sea surface and follows the sloping floor, as the water around it does.
The vertical velocity relative to a fixed datum would have to contain
these motions of the surface and the floor. KELP does not need it,
because they are already contained in \(H_p\).

Sea-surface height is read alongside the velocities and uses the same
method of temporal reconstruction (Appendix~\ref{app:timeint}). On sigma
meshes, recorded depths are measured from the instantaneous sea surface.
If sea-surface height is unavailable, KELP reports a warning and sets
\(\eta=0\).

Deterministic depths are clamped to the surface and local floor, with a
default 0.1~m bottom offset on sigma meshes. Scalar sampling applies the
same horizontal methods at the depths where the field is stored,
excluding non-finite neighbours. Backward tracking leaves the sign of
sampled values unchanged.

\subsection{Time integration and temporal reconstruction}\label{app:timeint}

With signed increment \(\delta t=s_{\mathrm{fb}}h\),
\(s_{\mathrm{fb}}=\pm1\), the coordinate increments are
\begin{equation}\label{eq:metric}
  \Delta\lambda = \frac{\delta t\, U}{M_\lambda(\phi)}, \qquad
  \Delta\phi = \frac{\delta t\, V}{M_\phi}, \qquad
  \Delta z = \delta t\, W ,
\end{equation}
where \(M_\phi=111\,120\)~m per degree and
\(M_\lambda=\max(M_\phi\cos\phi,1)\)~m per degree. After each update,
longitude is wrapped to \([-180^\circ,180^\circ)\). RK4 evaluates the
metric at the latitude of each stage. On sigma meshes, the fractional
depth \(s\) replaces \(z\), with \(\Delta s=\delta t\,W/H_p\)
(Appendix~\ref{app:vertical}).

The Dormand--Prince 5(4) scheme \citep{DormandPrince1980} advances each
particle through adaptive sub-steps within \(h\), using the usual
step-size control of embedded Runge--Kutta pairs and a position
tolerance given in metres (1~m by default).
This adaptive scheme is restricted to two dimensions. It uses velocities
at cell centroids, the plane fit over the cells around the nearest node
and either steady
or linearly reconstructed fields. It can be combined with constant
horizontal diffusion, but not with vertical diffusion or the edge walk.

Linear reconstruction between snapshots \(t_a\) and \(t_b\) is
\begin{equation}\label{eq:tlin}
  \mathbf{u}(t) = (1 - \alpha)\, \mathbf{u}_a + \alpha\, \mathbf{u}_b , \qquad
  \alpha = \frac{t - t_a}{t_b - t_a} ,
\end{equation}
with local interval \(\Delta T=t_b-t_a\). For smooth velocity, the field
reconstruction error is \(O(\Delta T^2)\). The same reconstruction is
applied to vertical velocity and sea-surface height.

By default, each tracking step spans one snapshot interval, so
\(h=\Delta T\). Within a step, Runge--Kutta stages still interpolate
between adjacent snapshots if the snapshot times are available. If a
multi-stage step extends past the last available snapshot, the current
field is held constant in time.

PCHIP reconstructs the velocity at a node from the four stored
snapshots nearest in time, two before and two after. The time derivative
of the velocity at each snapshot is estimated from the weighted harmonic
mean of the two neighbouring secants \citep{FritschButland1984}. It is
set to zero where the two secants have opposite signs or either is zero,
so that the reconstruction does not overshoot. Each velocity component
is reconstructed separately. The reconstructed field is therefore not
exactly divergence-free between snapshots, even if the stored fields
are. Reconstruction reverts to linear at the ends of the
record and where repeated forcing cycles meet. Slope limiting can reduce
the local order near extrema.

\subsection{Diffusion}\label{app:diffusion}

For constant horizontal diffusivity \(K_h\), each component receives
\begin{equation}\label{eq:rw}
  \Delta x_{\mathrm{s}} = \xi \sqrt{2 K_h\, |\delta t|} ,
\end{equation}
where \(\xi\) is a standard normal variate drawn independently for each
component. The displacement is mapped from the local tangent plane along
a great circle, using the radius \(R_{\mathrm{m}}=180M_\phi/\pi\)
consistent with the advection metric. Unequal components \(K_x,K_y\) add
a northward metric drift \((K_x-K_y)\tan\phi\,|\delta
t|/R_{\mathrm{m}}\). The latitude used to calculate this drift is
limited to \(\pm89.9^\circ\).

For spatially varying diffusivity, Euler--Maruyama adds
\(\partial_xK_x\,|\delta t|\) to the \(x\) increment \citep{Visser1997,
RossSharples2004}. Milstein also adds \(\tfrac12\partial_xK_x\,|\delta
t|(\xi^2-1)\) \citep{Grawe2012}. The \(y\) increment uses \(K_y\) and
\(\partial_yK_y\). The Milstein correction assumes that diffusivity is
diagonal and separable. It does not generally provide higher-order
accuracy for fields that vary in two dimensions.

Diffusivity is interpolated spatially by a plane fit and reconstructed
linearly in time, even when PCHIP is used for velocity. Gradients are
estimated by centred differences. The offsets follow the median
mesh-edge length and are bounded to 100~m--50~km. If an offset sample
lies outside the mesh, the centred estimate is replaced by one-sided
differences or the plane-fit gradient.

All integrators apply constant horizontal diffusion after the advective
step. With Euler, the spatially varying schemes combine advection and
diffusion in a step evaluated at the initial position. With RK2 or RK4,
optional splitting applies diffusion after advection, with the diffusion
terms evaluated at the initial or final position. The spatially varying
horizontal schemes are restricted to two dimensions.

Vertical diffusion uses
\begin{equation}\label{eq:az}
  \Delta z_{\mathrm{s}} = \frac{\partial A_z}{\partial z}\, |\delta t| + \xi \sqrt{2 A_z\, |\delta t|} ,
\end{equation}
where \(A_z\) and its gradient are evaluated on the same linear depth
segment in the column at the nearest node \citep{Visser1997}.
Displacements that cross the surface or floor are reflected at that
boundary. Variable \(A_z\) is available for global fixed-level output, while
the other input modules use a constant value. Vertical diffusion can
be combined with constant horizontal diffusion and with the fixed-step
integrators.

Random numbers are generated by a SplitMix64 chain \citep{Steele2014}
with the counter-based principle of \citet{Salmon2011}. Each draw is
keyed by the seed, realization, particle index in the full release,
global step and its use. Separate keys are used for horizontal
diffusion, vertical kicks and stranding decisions. Statistical and reproducibility checks
are summarized in Appendix~\ref{app:tests}.

\subsection{Stranding}\label{app:stranding}

Stranding combines criteria based on water depth, coastal cells and
stochastic beaching in the near-shore zone. A cell is classified as
coastal if one of its edges is not shared with another cell. After a
failed step, a particle can be reclassified as beached if its position
before the step lies in the near-shore zone. This zone includes water
shallower than the specified zone depth, whose default is twice the
stranding depth. Coastal cells are also included when that information
is available. Particles whose steps fail outside the zone remain
removed.

Resuspension can be disabled, triggered automatically by rewetting,
decided stochastically, or both. Each active mode requires rewetting
before the particle can move again. Hysteresis in the depth threshold
reduces repeated switching between beached and moving states. Automatic
release additionally requires a specified number of consecutive wet
steps.

\subsection{The rotated frame}\label{app:frame}

The optional rotation uses either the FESOM frame or specified Euler
angles. Node coordinates are rotated as points. Horizontal velocity
components are transformed as
\begin{equation}\label{eq:rot}
  u' = c\, u - s\, v, \qquad v' = s\, u + c\, v ,
\end{equation}
where \(c\) and \(s\) are the cosine and sine of the local rotation
angle. Vertical velocity, sea-surface height and scalars remain
unchanged. Anisotropic horizontal diffusivity is applied along the
tracking axes, with its components left unrotated. The FESOM-C input
module does not offer frame rotation and refuses a configuration that
requests it.

\section{Software details}\label{app:software}

\subsection{Input and memory}\label{app:io}

The input modules retrieve snapshots from NetCDF or Zarr as they are
needed. The HEALPix module can also access GRIB archives through reference files
that index their contents. Time-averaged records can be placed at the
midpoint of each averaging window. Experiments can be specified using
calendar dates, repeated forcing cycles and realizations shifted in
time.

The snapshots required for linear or cubic reconstruction are kept in
memory. Several consecutive snapshots can be retrieved together in a
single disk read. Single-precision storage halves the
memory used by velocity arrays. Large three-dimensional HEALPix fields
are loaded in blocks of levels. Processes running on the same node can
share snapshots through read-only memory maps.

\subsection{Extraction, particle sets and output}\label{app:output}

Regional extraction selects the triangular cells of the model mesh or
HEALPix pixels
inside a specified region. A halo, a rim of neighbouring cells around
the region, can be added. For
HEALPix, a single halo layer reproduces the dual mesh that a whole-mesh
build gives inside the region. Tests using a two-layer halo confirm
bit-for-bit agreement of trajectories for particles remaining within the
region.

Release files define particle positions, depths and optional time
offsets. A release can be repeated at fixed time intervals. For
parallel execution, a release file can be sorted by HEALPix pixel to
keep nearby particles together in memory. The file can then be divided
into contiguous subsets. When the release is not repeated, the resulting
Zarr trajectories are merged to restore the particle order of the
original release file. The
merged values match an unsplit run, including stochastic runs, provided
each subset keeps the same random seed and particle indices in the full
release.

Output records particle positions and status, together with any sampled
variables. Zarr output can be read during tracking. With buffered
writing enabled, NetCDF can be read between flushes, when the file is
closed. Restart files store particle state, time indices, stranding
counters and the coordinate frame. A run script uses these records to
resume the calculation through a chain of jobs.

\subsection{Compiled step routines}\label{app:kernels}

As described in Sect.~\ref{data-flow}, KELP either compiles the cell
search, interpolation and integration into one routine or runs them as
separate stages. The combined routine covers two-dimensional tracking,
three-dimensional Euler with the plane fit, and three-dimensional RK2
with node velocities interpolated from the vertices. Its availability
also depends on the chosen search and interpolation methods.
Three-dimensional sigma tracking, three-dimensional RK4, the edge walk
and the plane fit over edge-sharing neighbours run as separate stages.

In the stage-by-stage execution, each stage distributes work over
particles in parallel. For large contiguous fields, cubic and
double-precision linear temporal reconstructions are computed in
parallel over mesh points. Tests check that the combined and
stage-by-stage paths give the same trajectories for configurations
supported by both.

\subsection{Configuration}\label{app:config}

The experiment is specified in a YAML configuration. When the file is
loaded, scheme names and combinations of options are checked for
compatibility. Checks involving input fields are performed at startup or
when the fields are first used. Restart checks include verification of
the coordinate frame. The reference manual describes the available
options and the rules governing their combinations.

\subsection{Documentation, skills and tests}\label{app:docs}

A source map identifies the functions and classes described by each
topic document. It also records the code revision against which the
document was last checked. A checking tool reports functions or classes
that can no longer be found, along with source files modified since the
previous check. The skills were developed alongside the engine to
address problems documented during the experiment campaigns. During
development, the skills were refined using written test scenarios. Files
prepared by an agent were checked against lists drawn from the campaign
records. These development checks are separate from the controlled
evaluation in Sect.~\ref{sec:demo-agent}.

\label{app:tests}The test suite comprises about 2500 automated tests,
most of which use synthetic meshes and velocity fields. Those
requiring external data run when the necessary data are available. The
comparisons cover the combined and stage-by-stage routines, resumed and uninterrupted
runs, HEALPix regional and whole-mesh input, and runs split across
processes, including stochastic runs. Tests also verify that optional
features do not affect the results when disabled.

Stochastic tests verify that changing any part of the random key changes
the draws. They also check the expected Brownian displacement moments
and agreement between restarted and uninterrupted runs. Invariance to
thread count was confirmed in the runs comparing model-mesh and HEALPix fields (Sect.~\ref{sec:demo-healpix}).

\noappendix

\codeavailability{KELP is licensed under the BSD 3-Clause licence and available on GitHub at
\url{https://github.com/CliDyn/kelp} (last access: 13 November 2026). Version
1.0 described here is archived on Zenodo at
\url{https://doi.org/10.5281/zenodo.24200490} \citep{Kuznetsov2026}.
The archive includes the reference manual, with its chapter on agent skills,
and the test suite. It also includes scripts and configurations for running
and analysing all simulations in Sects.~\ref{sec:verification}--\ref{sec:discussion}.}

\dataavailability{The nextGEMS production simulation used in
Sect.~\ref{sec:demo-performance} is described by \citet{Segura2025}. Its
output is stored at DKRZ and is accessible with a DKRZ account. The
FESOM2 simulation on the CORE2 mesh used in
Sects.~\ref{sec:ver-reference}, \ref{sec:demo-healpix}
and~\ref{sec:demo-arctic} follows \citet{Danilov2017}. Its output is
stored at DKRZ and is available to DKRZ users. The HEALPix versions of
these fields (Sect.~\ref{sec:demo-healpix}) are not archived, because the
remapping scripts in the Zenodo archive rebuild them from the model
output. The wind-farm simulations of Sect.~\ref{sec:ver-reference} are
described by \citet{Androsov2026}, and their velocity output is stored at
the Alfred Wegener Institute. The configurations, release files, run and
processing scripts and the data behind the figures and tables of this
paper are in the same Zenodo archive as the code.}

\authorcontribution{All authors contributed to the conception and to the writing and revision of the manuscript, and approved the final version.}

\competinginterests{The contact author has declared that none of the
authors has any competing interests. }

\begin{acknowledgements}
We acknowledge the continuous support by the Alfred Wegener Institute, Helmholtz Centre for Polar and Marine Research. We also thank the Helmholtz Association with its (data) infrastructure investments as well as all direct and indirect funders (DFG, BMFTR, EU). This work is also supported by project S1: Diagnosis and Metrics in Climate Models of the Collaborative Research Centre TRR 181 ``Energy Transfer in Atmosphere and Ocean'', funded by the Deutsche Forschungsgemeinschaft (DFG, German Research Foundation, project no.\ 274762653). It is also supported by the EERIE project (Grant Agreement No 101081383), funded by the European Union, and by the TerraDT project (Digital Twin of Earth system for Cryosphere, Land surface and related interactions), which has received funding from the European Union's Horizon Europe research and innovation programme under Grant Agreement no.\ 101187992.
Views and opinions expressed are those of the authors only and do not necessarily reflect those of the European Union or the European Climate Infrastructure and Environment Executive Agency (CINEA). Neither the European Union nor the granting authority can be held responsible for them.
A.~Androsov acknowledges funding from the German Federal Ministry of Research, Technology and Space (BMFTR) through the projects CREATE (03F0910B) and CREATE~II (03F0979A), within the sustainMare research mission of the German Marine Research Alliance (DAM).

We developed KELP and carried out its experiments using coding agents
based on Anthropic and OpenAI models (Sect.~\ref{agent-layer}).
These tools also helped revise the authors' manuscript drafts under
their direction. The authors reviewed and edited the resulting text and
take full responsibility for the content.
\end{acknowledgements}

\bibliographystyle{copernicus}
\bibliography{kelp}

\end{document}